\documentclass[iop,apj,numberedappendix,twocolappendix]{emulateapj}

\usepackage{mathptmx} 
\usepackage[utf8]{inputenc}
\usepackage{amsmath}
\usepackage{amsfonts}
\usepackage{amssymb}
\usepackage{booktabs}
\usepackage{natbib}
\usepackage{graphicx}
\usepackage{aas_macros}
\usepackage[usenames,dvipsnames]{color}
\usepackage{xspace}
\usepackage{bm}
\usepackage{nicefrac}
\usepackage{cancel}
\usepackage{tabularx}

\definecolor{refkey}{rgb}{1,1,1}	
\definecolor{labelkey}{rgb}{1,0,0}	

\usepackage[normalem]{ulem} 
\usepackage{enumitem} 
\setlist{listparindent=\parindent,leftmargin=*}
\usepackage{svn-multi}

\definecolor{webbrown}{rgb}{.6,0,0}

\definecolor{webblue}{rgb}{0,0,0.6}

\definecolor{purple}{rgb}{0.5,0,.5}

\definecolor{webgreen}{rgb}{0,0.5,0}

  \newcommand{\cm}{\,{\rm cm}}

  \newcommand{\erg}{\,{\rm erg}}

  \newcommand{\kms}{\,{\rm km\,s^{-1}}}

  \newcommand{\K}{\,{\rm K}}
  \newcommand{\kpc}{\,{\rm kpc}}
  \newcommand{\p}{\,{\rm pc}}
  
  \newcommand{\Myr}{\,{\rm Myr}}
  \newcommand{\Gyr}{\,{\rm Gyr}}
  
  \newcommand{\mkG}{\,\mu{\rm G}}

  \newcommand{\Msol}{\,{\rm M_{\sun}}}
  \newcommand{\radm}{\,{\rm rad\,m^{-2}}}
  \newcommand{\s}{\,{\rm s}}
  \newcommand{\yr}{\,{\rm yr}}

  \newcommand{\lo}{l_{0}}
  \newcommand{\vect}[1]{\boldsymbol{#1}}
  \newcommand{\dd}{\mathrm{d}}

\usepackage[breaklinks,colorlinks,
  urlcolor=blue,citecolor=blue,linkcolor=blue,pdfborder={0 0 0.1}]{hyperref}
\hypersetup{
pdftitle={The supernova-regulated ISM -- V. Space- and time-correlations},
pdfauthor={
            J.~F.~Hollins (orcid.org/0000-0002-4435-4156),
            G.~R.~Sarson (orcid.org/0000-0001-6774-9372),
            A.~Shukurov (orcid.org/0000-0001-6200-4304),
            A.~Fletcher, 
            F.~A.~Gent, 
          }
}
\shorttitle{Supernova-regulated ISM -- V. Space- and time-correlations}
\shortauthors{J.~F.~Hollins, G.~R.~Sarson, A.~Shukurov, A.~Fletcher, F.~A.~Gent}

\begin{document}

\begin{abstract}
We apply correlation analysis to random fields in numerical simulations of the
supernova-driven interstellar  medium (ISM) with the magnetic field produced by 
dynamo action. We solve the thermo-magnetohydrodynamic (MHD) equations in a 
shearing, Cartesian box representing a local region of the ISM, subject to thermal 
and kinetic energy injection by supernova explosions, and parameterized 
optically-thin radiative cooling.
We consider the cold, warm\, and hot phases of the ISM separately; the 
analysis mostly considers the warm gas, which occupies the bulk of the 
domain. Various physical variables have different correlation 
lengths in the warm phase: $40\p$, $50\p$, and $60\p$ for random 
magnetic field, density, and velocity, respectively, in the midplane. 
The correlation time of the random velocity is comparable to the eddy turnover 
time, about $10^7\yr$, although it may be shorter in regions with higher star
formation rate.
The random magnetic field is anisotropic, with the standard deviations of 
its components $b_x/b_y/b_z$ having approximate ratios $0.5/0.6/0.6$ in the 
midplane. 
The anisotropy is attributed to the global velocity shear from galactic 
differential rotation, and locally inhomogeneous outflow to the galactic 
halo.
The correlation length of Faraday depth along the $z$-axis, $120\p$, is 
greater than for electron density, $60\text{--}90\p$, and vertical 
magnetic field, $60\p$. Such comparisons may be sensitive to the 
orientation of the line of sight.
Uncertainties of the structure functions of synchrotron intensity rapidly 
increase with the scale. This feature is hidden in power spectrum 
analysis, which can undermine the usefulness of power spectra for 
detailed studies of interstellar turbulence.
\end{abstract}

\title{Supernova-regulated ISM -- V. Space- and time-correlations}
\author{J.~F.~Hollins,\altaffilmark{1}
  G.~R.~Sarson,\altaffilmark{1} A.~Shukurov,\altaffilmark{1} A.~Fletcher,\altaffilmark{1}
  and F.~A.~Gent\altaffilmark{2}
}

\affil{$\sp{1}$ School of Mathematics and Statistics, Newcastle University,
Newcastle upon Tyne, NE1 7RU, UK. \\
$\sp{2}$ ReSoLVE Centre of Excellence, Department of Computer Science, Aalto 
University, PO Box 15400, FI-00076 Aalto, Finland}

\email{j.hollins@newcastle.ac.uk (JFH),
       graeme.sarson@newcastle.ac.uk (GRS),
       anvar.shukurov@newcastle.ac.uk (AS),
       andrew.fletcher@newcastle.ac.uk (AF),
       frederick.gent@aalto.fi (FAG)
       }


\section{Introduction}
\label{sect:introduction}
The interstellar medium (ISM) of a spiral galaxy is a complex, multiphase, random
system, driven by the input of thermal and kinetic energy from supernova (SN)
explosions and stellar winds \citep[e.g.,][]{MacLow:2004,Elmegreen:2004,Scalo:2004,
MacLow:2005,deAvillez:2005,Federrath:2010,Hill:2012}.
Its statistical analysis, including that of interstellar turbulence, is complicated 
by the multi-phase structure, where the diversity of physical processes predominant 
in different phases causes strong inhomogeneity. Furthermore, interstellar 
turbulence is transonic or supersonic \citep{Bykov:1987,Vazquez-Semadeni:2015}. 
The compressibility and abundance of random shock waves lead to spatial and temporal
intermittency of the random velocity and magnetic fields and of the density 
fluctuations. Dynamo action adds further complexity by producing intermittent random
magnetic fields \citep{Wilkin:2007}.

Observational studies of such an inhomogeneous, complex random system are severely
limited by the fact that observable quantities are integrals along the 
line-of-sight, so that many physically significant statistical features become 
hidden. When observed at a low resolution, the interstellar medium can be 
satisfactorily described in terms of Gaussian random fields, but recent observations
have revealed a plethora of density structures in neutral hydrogen, mostly planar or 
filamentary \citep[][and references therein]{Heiles:2003,Makarenko:2015,Wang:2016}. 
Statistical analysis of such random fields cannot be restricted to the standard 
tools of the theory of Gaussian random functions (and related ones, such as 
log-normal and $\chi^2$ functions), where the probability distribution and 
second-order correlation functions provide a complete description. However,
correlation analysis remains an important first step, where the form of the
correlation function, the correlation length (or time) and the mean-square 
variations of a variable are the most important quantities explored.

There are numerous and diverse estimates of the integral (correlation) scale of
interstellar turbulence $\lo$ \citep[see also][]{Haverkorn:2013}.
The autocorrelation function of the line-of-sight \ion{H}{1} cloud velocities 
obtained in the Milky Way by \citet{Kaplan:1966} leads to $\lo\simeq80\p$.
\citet{Lazaryan:1990} found $\lo\approx50\p$ from the fluctuations in synchrotron
intensity. \citet{Ohno:1993} used differences in Faraday rotation between 
neighboring pulsars to obtain $10<\lo<100\p$. \citet{Minter:1996} yielded
$\lo\simeq4\p$ from the structure functions of the variations in the Faraday 
rotation and emission measures across extended extragalactic radio sources.  
Structure functions of the Faraday rotation of extragalactic sources 
\citep{Haverkorn:2004, Haverkorn:2006, Haverkorn:2008} and their degree of
depolarization \citep{Haverkorn:2008} give $\lo \simeq 1 \p$ in the Milky Way's 
spiral arms. $\lo<20\p$ was found by an analysis of low-frequency synchrotron
intensity fluctuations from a large region of the Galactic disk by
\citet{Iacobelli:2013}.
In the Large Magellanic Cloud, the structure function of the Faraday rotation of 
more distant sources gave $\lo \simeq 90 \p$ \citep{Gaensler:2005}.
In the galaxy M51  \citet{Fletcher:2011} obtained $\lo\simeq50\p$ from the 
depolarization of diffuse emission, whilst \citet{Houde:2013} found $\lo\simeq65\p$ 
from the dispersion of radio polarization angles.
These estimates are strikingly different, perhaps not surprisingly. They have been
obtained from diverse tracers, and it is not surprising that the correlation length 
of the gas velocities, Faraday rotation measure and synchrotron fluctuations
differ (the latter being nonlinear functions of the fluctuating quantities).
A relation between the correlation length of the product of random functions and 
those of the multipliers depends on their detailed statistical properties 
\citep[e.g., \S6.2 in][]{Stepanov:2014}. Our aim here is to clarify this relation.
This would be difficult to do with observational data, at least at present.

Interpretations of observations of polarized synchrotron emission and its Faraday
rotation suggest that a significant fraction of the polarization may be due to
anisotropy of the random magnetic field. 
The correlation between the mean Faraday rotation and its standard deviation
along the Galactic disc, found by \citet{Brown:2001}, was the earliest indication 
of an anisotropic random field. 
Subsequent models of various components of Milky Way emission along the 
Galactic disk \citep{Jaffe:2010, Jaffe:2011, Jaffe:2013} and across the entire
sky \citep{Jansson:2012a, Jansson:2012b} required the inclusion of an
anisotropic random magnetic field in order to fit the observations.
In other galaxies, modeling of pre- and post-shock polarized emission in the
barred galaxies NGC1097 and NGC1365 \citep{Beck:2005a} and the spiral galaxy 
M51 \citep{Fletcher:2011}, the dispersion of polarization angles in M51
\citep{Houde:2013}, comparison of the observed polarized emission and Faraday 
rotation in M33 \citep{Stepanov:2014}, and modeling depolarization in M51
\citep{Shneider:2014}, have all indicated the presence of anisotropic random fields.
Extracting the degree of anisotropy from the observations, though, is
difficult.

In M51, \citet{Fletcher:2011} estimate that the ratio of the standard deviations of
the random magnetic field components in orthogonal directions is $\sigma_x/\sigma_y\simeq 2$ and \citet{Houde:2013} obtained a ratio of correlation
lengths along and perpendicular to the local mean-field direction of
$l_\parallel/l_\perp \simeq 1.8$.
As with observational estimates for $\lo$, it is appropriate to carefully examine 
the possible anisotropy of the random magnetic field.

Simulations of the SN-regulated ISM have become sufficiently realistic to treat them
as numerical experiments. It is then natural to use sufficiently realistic numerical
models to address these questions before the more difficult observational 
exploration.
We use such simulations, as detailed in \citet{Gent:2012} and
\citet{Gent:2013a,Gent:2013b}, which have non-trivial magnetic fields generated by
dynamo action, to clarify the correlation (and other statistical) properties of the
multi-phase ISM. In particular, we compare the autocorrelation and cross-correlation
functions of the random (i.e., small-scale; see \S\ref{sect:averaging})
velocity and magnetic fields and density fluctuations, as well as the Faraday depth
and synchrotron intensity.

However complex, the simulations of the ISM can hardly be considered as trustworthy
representations of the ISM in its whole complexity. Therefore, the goal of our
analysis is not to achieve quantitative agreement with observations in every detail
(although the general agreement is quite remarkable) but rather to identify those
physical processes that shape the simulated ISM and are likely to be important in
reality.

Turbulent flows are often represented in spectral space, in terms of the Fourier
transforms of the physical variables. Such transforms are straightforward in 
infinite or periodic spaces. However, simulations of the ISM are performed in
relatively small domains, only containing of order one thousand correlation cells, 
not simply-periodic because of the open (or similar) boundary conditions at the top
and bottom of the domain, and statistically inhomogeneous because of the
stratification \citep[e.g.][]{Korpi:1999a,Korpi:1999b,Gent:2013a}.
Furthermore, it is difficult to estimate reliably the statistical uncertainty of the
Fourier transforms.

We therefore proceed via correlation analysis \citep[e.g.,][]{Monin:1975}.
For most of the work, we assume local isotropy in the horizontal ($xy$) plane; this
assumption is assessed in \S\ref{sect:aniso}.

Correlation lengths obtained from comprehensive numerical simulations of the 
multi-phase ISM exhibit less diversity than the observational results.
\citet{Joung:2006} obtain a gas density spectrum with a peak at $20\p$, 
whereas most kinetic energy is contained at scales $20\text{--}40\p$. 
\citet{Gent:2013a} calculate $\lo=100\p$ for the random velocity field in the
mid-plane of the galaxy, also from hydrodynamic simulations.
In the MHD simulations of \citet{deAvillez:2007}, $\lo=70\p$ for the random velocity
field. This scale fluctuates strongly with time. 
From correlation analysis of the vertical component of random velocity,
\citet{Korpi:1999a} obtained an estimate of $\lo=30\p$ for the warm gas at all
heights, whereas in the hot gas $\lo$ increases from $20\p$ in the mid-plane to 
$60\p$ at $|z|=150 \p$.

The paper is organized as follows. The simulations of the SN-driven ISM and 
averaging procedure used in our analysis are presented in \S\ref{sect:model}. 
The spatial correlations of the random magnetic field, density and velocity are
discussed in \S\ref{sect:spatial}, whereas time correlations are the subject of
\S\ref{sect:time_correlation}.
The anisotropy of the random magnetic field in the simulated ISM is estimated and
interpreted in \S\ref{sect:aniso}.
The autocorrelation functions of such observable quantities as the Faraday depth and synchrotron intensity are obtained and discussed in \S\ref{sect:observables}. 
Our results are summarized in \S\ref{sect:summary}. Appendix 
\ref{sect:domain_comparisons} presents a comparison with the results obtained in 
a larger computational domain.

\section[]{Simulations of the multi-phase ISM}
\label{sect:model}
We use our earlier simulations of the ISM based on the
\textsc{Pencil Code} (\url{https://github.com/pencil-code}),
using its ISM modules that implement SN energy injection and radiative
cooling, and handle shocks produced in a supersonic flow,
described in detail by \citet{Gent:2012} and \citet{Gent:2013a}.

The simulations involve solving the full, compressible, non-ideal MHD equations with
parameters generally typical of the solar neighborhood in a three-dimensional local
Cartesian, shearing box with radial ($x$) and azimuthal ($y$) extents of 
$L_{x} = L_{y} = 1.024 \kpc$ and vertical ($z$) extent $L_{z} = 1.086 \kpc$ 
on either side of the mid-plane at $z = 0$.

Our numerical resolution is $\Delta x = \Delta y = \Delta z = 4 \p$, using $256$
grid points in $x$ and $y$ and $544$ in $z$. \citet{Gent:2013a} demonstrate that 
this resolution is sufficient to reproduce the known solutions for expanding SN
remnants in the Sedov--Taylor and momentum-conserving phases. 

Details of the numerical implementation and its comparison with other 
similar simulations can be found in Section~\ref{sect:parameters}.

The basic equations are mass conservation, the Navier--Stokes equation, the heat
equation, and the induction equation, solved for mass density $\rho$, velocity
$\boldsymbol{u}$, specific entropy $s$, and magnetic vector potential 
$\boldsymbol{A}$ (such that $\boldsymbol{B}=\nabla\times\boldsymbol{A}$).

The Navier--Stokes equation includes a fixed vertical gravity force that includes
contributions from the stellar disk and spherical dark halo. The initial state is an
approximate hydrostatic equilibrium. The Galactic differential rotation is modelled 
by a background shear flow $\boldsymbol{U} = (0,-q \Omega x, 0)$, where $q$ is the
shear parameter and $\Omega$ is the Galactic angular velocity.
Here we use $q = \boldsymbol{+1}$, as in a flat rotation curve, and 
$\Omega = 25 \kms \kpc^{-1}$, as in the Solar neighborhood.

The velocity $\boldsymbol{u}$ is the perturbation velocity in the rotating frame, 
that remains after the subtraction of the background shear flow from the total
velocity. However, it still contains a large-scale vertical component due to an
 utflow driven by the SN activity.

Both Type II and Type I SNe are included in the simulations. These differ in their
vertical distribution and frequency only. The frequencies used correspond to those 
in the Solar neighborhood. We introduce Type II SNe at a mean rate, per unit surface
area, of $\nu_{\rm II} = 25 \kpc^{-2} \Myr^{-1}$. Type I SNe have a mean rate, per
unit surface area, of $\nu_{\rm I} = 4 \kpc^{-2} \Myr^{-1}$.

The SN sites are distributed randomly in the horizontal planes. Their vertical
positions have Gaussian distributions with scale heights of 
$h_{\rm II} = 0.09 \kpc$ and $h_{\rm I} = 0.325 \kpc$ for SNII and SNI, 
respectively. No spatial clustering of the SNe is included. The thermal energy 
injected with each SN is $0.5 \times 10^{51} \erg$. Injected velocity and the 
uneven density within each explosion site randomly adds kinetic energy with mean 
$0.4 \times 10^{51} \erg$.

We include radiative cooling with a parameterized cooling function. For 
$T < 10^{5} \K$, we adopt a power-law fit to the `standard equilibrium' 
pressure--density curve of \citet[][]{Wolfire:1995}, as given in 
\citet[][]{Sanchez-Salcedo:2002}. For $T > 10^{5} \K$, we use the cooling function 
of \citet[][]{Sarazin:1987}. This cooling allows the ISM to separate into distinct 
hot, warm and cold phases identifiable as peaks in the joint probability 
distribution of gas in density and temperature.

Photoelectric heating is also included as in \citet[][]{Wolfire:1995}. The heating
decreases with $|z|$ on a length scale comparable to the scale height of the disk 
near the Sun.

Shock-capturing kinetic, thermal and magnetic diffusivities (in addition to constant
small background diffusivities), are included to resolve shock discontinuities and
maintain numerical stability in the Navier--Stokes, heat and induction equations.

Periodic boundary conditions are used in $y$, and sheared-periodic boundary 
conditions in $x$ (considered in more detail in \S\ref{sect:shear}).
Open boundary conditions, permitting outflow and inflow, are used at the vertical
($z$) boundaries.
See \citet[][]{Gent:2012} and \citet[][]{Gent:2013a,Gent:2013b} for further details 
on the boundary conditions used and on the other implementations described above.

Starting with a weak initial azimuthal magnetic field at the mid-plane,
this system is susceptible to the dynamo instability. Dynamo action can be 
identified \citep{Gent:2013a} with exponential field growth saturating after
$1.4\Gyr$, at root mean square field strengths of order
$2.5\mkG$, comparable to observational estimates for the solar
neighbourhood.

Our analysis is based on 12 snapshots of the computational volume in the range 
$1.4 \leq t \leq 1.675 \Gyr$, by which time the system, including the large-scale
magnetic field, has reached a statistically steady state. The interval between the
snapshots, $25 \Myr$, is significantly longer than the correlation time of the 
random flow (see \S\ref{sect:time_correlation}), and is sufficient for the snapshots
to be considered statistically independent.

To test the influence of shear rate on the correlations, we also analyse data from a
model with twice the rotation rate, as discussed in \citet{Gent:2013b}. We use $12$
snapshots in the range $1.4 \leq t \leq 1.675 \Gyr$, again with a separation of 
$25 \Myr$, with the magnetic field saturated as for the main run. Any notable
differences between the results for the different models will be reported 
throughout the text.

\subsection{Parameters of the numerical model}
\label{sect:parameters}
	
    The model discussed here aims to reproduce the statistical properties of 
    the random ISM. With the integral scale of random fluctuations in various
	physical variables of order $50\p$ (see below), 
	the computational domain that
	we use contains about $400$ correlation cells, providing 
	sufficient statistics to obtain useful results. Other simulations of comparable
	physical content \citep{Hill:2012,Bendre:2015} have computational boxes of a
	similar horizontal size of $0.8\text{--}1\kpc$. 
	Physically distinct objects of the next largest scale are 
	superbubbles, of order $0.5\text{--}1\kpc$ in size,
	and OB associations and spiral arms whose scale is of order
	$1\text{--}3\kpc$; modelling these phenomena
	would require significantly larger computational domains (and the next generation
	of computational models) although some of their features can be captured with
	existing models \citep[e.g.,][]{Shukurov:2004,deAvillez:2007}.
	
    The size of the superbubbles produced by SNe clustering are comparable to 
    the horizontal size of the computational domain. As a result, we neglect
    the clustering of SNe, although it would not be difficult to include it. 
    Simulations in a domain of a significantly larger size are required 
	to capture the effects of the SN clustering.
	\citet{deAvillez:2007} include SN clustering in their simulations and obtain
	the correlation scale of the random flows of $\boldsymbol{75}\p$, 
	comparable to those obtained below. In order to fully understand the effects
	of clustering, simulations without clustering must first be understood, which
	is the purpose of the current work.
	
    The vertical size of the domain is largely controlled by its horizontal 
	size. 
	A vertical extent of $1\kpc$ is insufficient to capture 
    fountain flows and model the temperature distribution in the halo, which require
    heights of greater than $5\kpc$ \citep[see][]{Hill:2012}.
    However, our simulations are able to fulfil our 
    purpose of capturing the physics of the ISM near the midplane, excluding 
    fountain flows, without any effects introduced via periodic boundary conditions.
	As argued by \citet{Gent:2013a}, periodic boundary conditions in the
	horizontal planes affect the outflow speed significantly at altitudes exceeding
	the horizontal extent of the region. Furthermore, the diameter of supernova 
	shells increases to $0.4\text{--}0.6\kpc$ at 
	$|z|\simeq1\kpc$. Therefore, results obtained at 
	$|z|\gtrsim1\kpc$ in a computational box of 
	$1\times1\kpc^2$ horizontally may be questionable. 
	Results from recent simulations performed in computational boxes taller than 
	$1\kpc$ are mostly reported only within 
	a few kiloparsecs from the midplane \citep[e.g.,][]{Hill:2012}. 
    The domain used in our simulations includes two scale heights of the warm 
    neutral gas.
	
    With the range of $|z|$ limited to $1\kpc$ 
    in our simulations, we have paid special effort to ensuring that the boundary
    conditions at the top and bottom boundaries do not introduce any apparent 
    artefacts into numerical solutions, such as a boundary layer with a strong
    gradient in any of the physical variables \citep[Appendix~C of][]{Gent:2013a}. 
	The limited vertical extent of the box is the main limitation
	of our model, but it can only be increased together with its horizontal size.
	Appendix~\ref{sect:domain_comparisons} presents results obtained with
	a slightly larger domain (three times the volume of the domain
	used for the main computations). We conclude that the results reported here
	are not affected by the change in the simulated volume.
	
    The mass loss rate through the top and bottom boundaries is about 
	$10^{-3}\Msol\yr^{-1}$,	 
	so $10^{6}\Msol$ is lost in $1\Gyr$,
	as compared to the total gas mass of 
	$10^{7}\Msol$ in the computational domain.
	This mass loss would correspond to a total mass loss rate of 
    $1\Msol\yr^{-1}$
    for a galactic disk of radius $15\kpc$,
    assuming the Galaxy is in a steady state. 
    Our open boundary conditions allow for inflow as well as outflow
    (albeit in a rather ad hoc way),
    which mitigates mass loss through the boundaries.
    The mass loss, albeit only modest, was compensated by a continuous mass 	
	replenishment (in proportion to the local gas density, for minimal impact 
	on the dynamics) to maintain an approximately constant gas mass throughout 
	the simulations.
	
    The numerical resolution of $4\p$ that we use has been 
    carefully selected to reproduce accurately the known expansion laws and 
    approximate internal structure of 
	an isolated supernova remnant, subject to radiative cooling 
    processes, until its radial shell expansion slows to match the ambient 
    adiabatic speed of sound \citep[Appendix~B of][]{Gent:2013a}. Thus, we are
    confident that our simulations model reliably the associated energy injection 
    into the diffuse ISM. Indeed, the intensity of random flows, of order 
	$10\kms$ in the warm gas and higher in the hot phase,
	is in full agreement with both observations and simulations at a higher
	resolution. This is also true of the scales of the random flows, fractional 
	volumes of the ISM phases and other aspects of the modelled ISM. 
	We have adjusted thermal conductivity so as to ensure that any
	structures produced by thermal instability are 
	fully resolved at the $4\p$ resolution. 
	Comparable simulations of \citet{deAvillez:2007,deAvillez:2012a} have an
	adaptive mesh with the finest separation of $1.25\p$, whereas
	\citet{Hill:2012} have a resolution of $2\p$, both representing 
	an arguably modest improvement. 
	We were unable to identify any differences in the relevant results of all these
	simulations that might be a consequence of the difference in numerical resolution.
	We also note that adaptive mesh refinement makes it difficult to include
	differential rotation, a crucial element of realistic modelling of magnetic
	field.

    Self-gravity is ignored in our simulations since we do not attempt to model
    the very cold molecular gas which is significantly affected by self-gravity. 
    Simulations with higher resolution would be required to model the higher 
    densities and the associated cooling rates.

\subsection{The multi-phase structure}
\label{sect:multiphase}
The numerical model exhibits three distinct states of the gas corresponding to local
maxima in the probability distribution function (PDF) of the specific entropy $s$.
Gas parameters in those states are similar to the three main phases of the ISM.
Following \citet{Gent:2013a}, the cold phase is defined 
as that having $s \leq 3.7 \times 10^{8} \erg \K^{-1}$,
the hot phase has $s \geq 23.2 \times 10^{8} \erg \K^{-1}$,
with the warm phase in between. The three phases have very different physical
properties, including the random velocity and magnetic fields, as well as differing 
in their mean temperature and density.
Therefore, our analysis is carried out for each phase separately.
For this purpose, only grid points corresponding to a given phase are retained in 
the data cubes containing each physical variable, with the other points masked out.
This allows us to do the averaging required in the computation of the structure
functions over disjoint regions in the physical space.

\begin{figure*}
\centering
\includegraphics[width=0.47\textwidth]{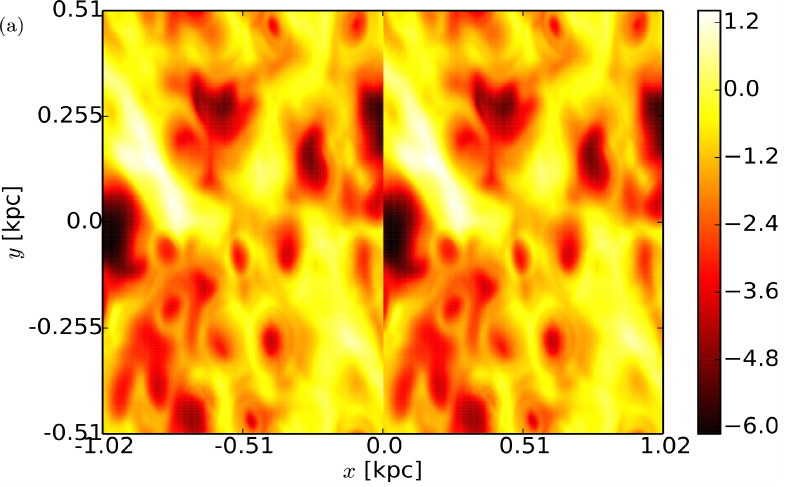}
\hfill
\includegraphics[width=0.47\textwidth]{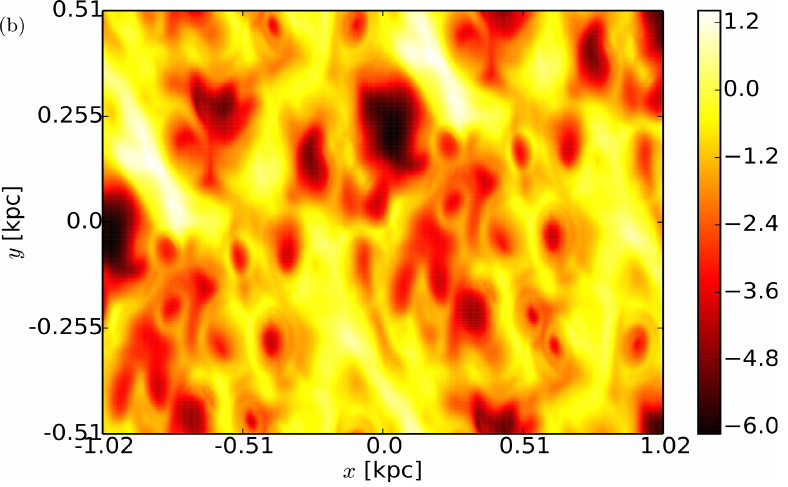}
\caption{Aligned domains of logarithm of gas number density (${\log n}$), at $z=2 \p, \, t=1.55 \Gyr$; 
(a) before and (b) after shifting the right-hand domain by $\delta y$ to account for the shearing boundary. 
The boundary between the two
copies of the computational domain is here located at $x=0$.
\label{fig:Figure1}
}
\end{figure*}

\subsection{Averaging procedure}
\label{sect:averaging}
Our analysis is conducted for the moduli of the random magnetic and velocity fields and the gas density fluctuations, denoted $b$, $u'$ and $n'$, respectively.
The random velocity $\vect{u}'$ should be carefully distinguished from the velocity
perturbation $\vect{u}$, defined as the deviation from the background large-scale
shear flow, since the latter contains a systematic vertical velocity.

Since the mean vertical velocity and the large-scale magnetic field are not 
necessarily uniform across any horizontal plane, we do not use horizontal averages 
to define the mean magnetic field as is often done in the literature, but instead 
follow \citet{Gent:2013b} and use Gaussian smoothing, within the framework of 
\citet{Germano:1992}.
The mean (large-scale) component of a random field $f$, averaged over a scale 
$\ell$ and denoted $\langle f\rangle_\ell$, is defined by a convolution with a
Gaussian kernel $G_{\ell}(\boldsymbol{x})$,
\begin{align}
\langle f \rangle_{\ell}(\boldsymbol{x}) = \int_{V} f(\boldsymbol{x}')\,G_{\ell}(\boldsymbol{x}-\boldsymbol{x}') \, \mathrm{d}^{3}\boldsymbol{x}', \notag \\
G_{\ell}(\boldsymbol{x}) = (2 \pi \ell^{2})^{-3/2}\exp[-\boldsymbol{x}^{2}/(2 \ell^{2})] \, ,
\label{eq:gauss}
\end{align}
where integration is extended to the volume occupied by a given ISM phase or the 
total volume as appropriate.
The random velocity is then 
$\boldsymbol{u}' = \boldsymbol{u} - \langle \boldsymbol{u} \rangle_{\ell}$
and similarly for the magnetic field,
$\boldsymbol{b} = \boldsymbol{B} - \langle \boldsymbol{B} \rangle_{\ell}$
and the gas number density, $n'=n-\langle n \rangle_{\ell}$.
Following \citet{Gent:2013b}, we use $\ell=50\p$.

As discussed by \citet{Gent:2013b}, a significant fraction of the energy in the 
random field remains at length scales greater than $\ell$. To clarify the 
consequences of this, consider averaging a random field $f(\boldsymbol{x})$ in wave
number ($\boldsymbol{k}$) space, denoting $\widehat{f}(\boldsymbol{k})$ the Fourier transform of $f(\boldsymbol{x})$.
By the convolution theorem, the mean field $\langle f\rangle_\ell(\boldsymbol{x})$ 
has the Fourier transform 
$\langle \widehat{f}\rangle_\ell(\boldsymbol{k})=\widehat{f}(\boldsymbol{k})\,\widehat{G}_{\ell}(\boldsymbol{k})$,
where $\widehat{G}_{\ell}(\boldsymbol{k})$ is the transform of the smoothing kernel.
For the Gaussian kernel $G_\ell(\boldsymbol{x})$, we have
$\widehat{G}_{\ell}(\boldsymbol{k})=\exp(-\ell^2 \boldsymbol{k}^2/2)$, so 
$\langle\widehat{f}\rangle_\ell(\boldsymbol{k})=
\exp(-\ell^2 \boldsymbol{k}^2/2) \widehat{f}(\boldsymbol{k})$.
Thus, for variations with wavenumber $k$ (and wavelength $\lambda=2\pi/k$),
a fraction $\exp(-\ell^2 \boldsymbol{k}^2)$ of the original field energy is
interpreted as that of the mean field, and the remainder goes into the random field.
This fraction is half -- i.e., the field energy is equally split between mean and
random fields -- at the wave number $k_{\rm eq}=\sqrt{\ln 2}/\ell$, or
the wavelength $\lambda_\mathrm{eq}=2\pi\ell/\sqrt{\ln2}\approx7.5\ell$. 
Thus, with $\ell = 50  \p$, the field energy is equally split at the wavelength
$\lambda_{\rm eq}\approx 380\p$ between the mean and random parts. 
Variations with wavelength $\lambda < 380  \p$ go predominantly into the random 
field, and increasingly so as $\lambda$ decreases; for features with 
$\lambda = 50 \p$, only a fraction $\exp(-4\pi^2) \approx 10^{-17}$ of the energy 
goes into the mean field.

\subsection{The structure and correlation functions}
\label{sect:functions}
We start the calculations with the second-order structure functions $D(l)$,
which are more robust than the correlation functions, $C(l)$, with respect to errors
\citep[\S13.1 in][]{Monin:1975}:
\begin{equation}
D(l) = \langle [f(\boldsymbol{x}+\boldsymbol{l})-f(\boldsymbol{x})]^{2} \rangle_l,
\label{eq:structure}
\end{equation}
where $\boldsymbol{x}$ a given position in the $(x,y)$-plane and $\boldsymbol{l}$ a
horizontal offset with $l = | \boldsymbol{l} |$. Analysis is restricted to
horizontal planes with no offsets in the $z$-direction, because of the 
stratification in $z$.

Since we are dealing with periodic (or sheared periodic) functions in $x$ and $y$, 
the maximum offsets we can consider in the $x$ and $y$ directions are half the 
domain sizes in each direction. 
Hence, we consider offsets in the range 
$0 \leq l_{x} \leq L_{x}/2, \, 0 \leq l_{y} \leq L_{y}/2$.
Using $D(l)$, the autocorrelation function $C(l)$ is obtained as
\begin{equation}
C(l) = 1 - \frac{D(l)}{2\sigma^{2}},
\label{eq:correlation}
\end{equation}
where $2\sigma^{2}$ is the value of $D(l)$ at which the random function 
$f(\vect{x})$ is no longer correlated, and $\sigma$ is the dispersion 
(r.m.s.\ value) of $f(\vect{x})$.
The choice of $2\sigma^{2}$ in a finite domain is not always obvious (see below). 
In terms of $C(l)$, the correlation length $l_{0}$ is defined as
\begin{equation}
l_{0} = \int_0^\infty C(l) \, \mathrm{d}l.
\label{eq:length}
\end{equation}
The magnitude of the implied correlation length is very sensitive to the range of
integration and to the behaviour of the correlation function at large $l$. An
exponentially small tail in $C(l)$ can make a significant contribution to $\lo$.

To address this problem, we fit the structure functions obtained from
equation~\eqref{eq:structure} to one of the following analytic forms 
(as discussed below), thereby obtaining estimates of $\sigma^{2}$ and 
$L_0$ (and hence $l_0$):
\begin{align}
D(l) &= 2 \sigma^{2} \left[1 - \exp \left(- \frac{l}{L_0} \right) \right], 
\qquad l_0 = L_0,
\label{eq:l_fit1}\\
D(l) &= 2 \sigma^{2} \left[1 - \exp \left(- \frac{l^{2}}{2 L_{0}^{2}} \right) \right],
\qquad l_0 = \sqrt{\frac{\pi}{2}}\, L_0.
\label{eq:l_fit2}
\end{align}
Since the governing equations contain second-order derivatives in spatial 
coordinates, the spatial variations must be smooth random functions of position, 
so that $\mathrm{d}C/\mathrm{d}l=0$ at $l=0$ for spatial correlations. 
However, the fact that only the first time derivatives appear in the governing
equations implies that the time variations only needs to be continuous, 
so that $\mathrm{d}C/\mathrm{d}\tau\neq0$ for $\tau=0$ may be expected 
for time correlations (as considered in \S\ref{sect:time_correlation}),
with $\tau$ the time lag.

We indeed observe this difference in the computed structure and correlation 
functions, and use form in equation~\eqref{eq:l_fit1} for time correlations and
equation~\eqref{eq:l_fit2} for spatial correlations.
Some of the spatial autocorrelation functions discussed below (most notably those 
for the density fluctuations) exhibit an oscillatory behavior; in such cases,
equation~\eqref{eq:l_fit2} is augmented to
\begin{equation}
D(l)=2\sigma^2 \left[1-\exp\left(-\frac{l^2}{2L_0^2}\right)\cos (k l)\right],
\label{eq:cf_fit}
\end{equation}
with $k = a l + b$, where $a$ and $b$ are two additional parameters determined by 
the zeros in the correlation function.

The correlation lengths $l_0$ are presented in Table~\ref{tab:Table1}.
To confirm the importance of using fitted correlation functions, we also present 
in this table the correlation lengths $\tilde{l}_0$ obtained by integration of the
directly calculated $C(l)$, over the range $0\le l \le 500\p$.
The values differ by up to a factor of $2$, with the differences being greatest for
density fluctuations (where the form in equation~\eqref{eq:cf_fit} was used);
the agreement for random magnetic field and velocity (where the form in
equation~\eqref{eq:l_fit2} was used) is closer.

To improve the reliability of our statistics, the averaging involved in the
calculation of the structure functions is performed over 26 grid planes within 
layers at $|z|\leq50\p$, $|z-0.4\kpc|\leq50\p$ and $|z+0.4\kpc|\leq50\p$ for each
snapshot, and then the structure functions are further averaged over the snapshots.
The uncertainty of the resulting values of the structure functions is rather small
(of order $10^{-3}$ in terms of the relative error) because of the large number of
data-point pairs available even at large values of $l$.
The structure and correlation functions in figures below are shown with error bars
representing \textit{not} their uncertainty but the standard deviation of the
individual measurements around the mean.

\begin{table*}
\centering
\caption{
The root-mean-square (rms) values of the fluctuations, their magnitude relative to the mean and correlation lengths of the fluctuations
in gas density, speed and magnetic field for the warm gas at
the mid-plane $z=0$ and at $|z|=400\p$.
\label{tab:Table1}}
\begin{tabular}{@{}l cccc @{}c cccc  @{}c cccc@{}}%
\hline\hline
$|z|$         &\multicolumn{4}{c}{Density fluctuations}                  & &\multicolumn{4}{c}{Random speed}                   & &\multicolumn{4}{c}{Random magnetic field}               \\
               \cline{2-5}                                                  \cline{7-10}                                          \cline{12-15}                                           \\[-8pt]
              &rms               &rms           &$l_0$    &$\tilde{l}_0$ & &rms           &rms        &$l_0$    &$\tilde{l}_0$ & &rms             &rms           &$l_0$    &$\tilde{l}_0$ \\
$[\text{pc}]$ &[cm$^{-3}$]       &relative      &[pc]     &[pc]          & &[km/s]        &relative   &[pc]     &[pc]          & &$[\mu$G]        &relative      &[pc]     &[pc]          \\
\hline
0             &$0.306\pm0.001$   &$0.49\pm0.03$ &$53\pm5$ &$24\pm1$      & &$8.10\pm0.03$   &$1.0\pm0.1$ &$60\pm3$ &$50\pm1$     & &$0.587\pm0.002$ &$0.58\pm0.04$ &$44\pm2$ &$41\pm1$      \\
400           &$0.0604\pm0.0001$ &$0.49\pm0.03$ &$37\pm2$ &$27\pm2$      & &$2.84\pm0.01$ &$0.49\pm0.11$ &$87\pm3$ &$81\pm1$     & &$0.483\pm0.001$ &$0.39\pm0.05$   &$64\pm2$ &$55\pm1$      \\
\hline
\end{tabular}
\tablecomments{Two values of the correlation lengths are provided,
$l_0$ obtained from a fitted form of the structure function as described in \S\ref{sect:functions}, and $\tilde{l}_0$ derived by integrating the calculated correlation function within the available range,
$0\leq l\leq 500\p$; the difference demonstrates how important is the fitting to obtain a reliable estimate of $\lo$.}
\end{table*}

\begin{figure*}
\centering
\includegraphics[width=0.47\textwidth]{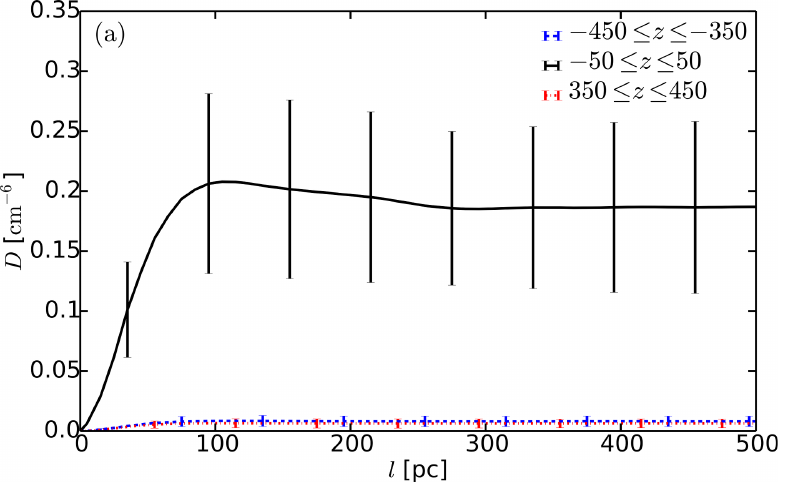}
\hfill
\includegraphics[width=0.47\textwidth]{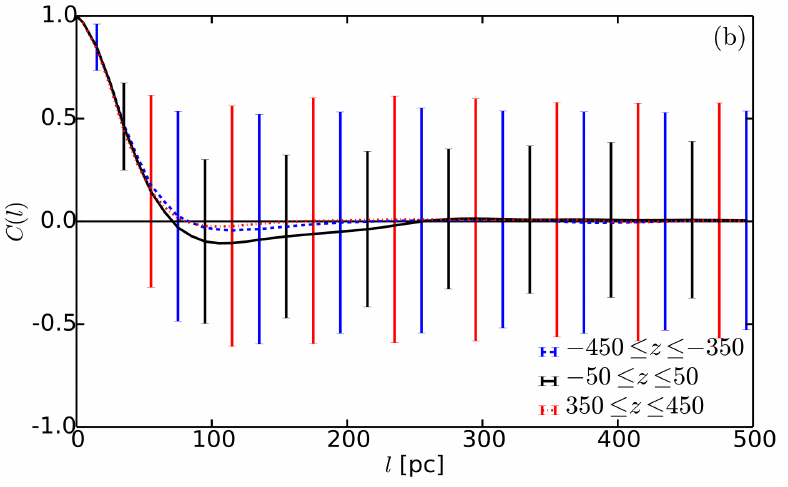}
\caption{(a) The structure function $D(l)$ and (b) correlation functions $C(l)$ for density fluctuations in the warm gas,
averaged about $z = -400 \p$ (blue, dashed), $z = 0 \p$ (black, solid), and $z = 400 \p$ (red, dash-dotted). 
The error bars denote the standard deviation of the individual measurements around their mean values, rather than
the error of the mean value, as discussed in \S{\ref{sect:functions}}.
For clarity, the error bars are only shown for every sixth bin in $l$, and 
staggered between the different curves.
\label{fig:Figure2}
}
\end{figure*}

\subsection{Accounting for shearing boundaries}
\label{sect:shear}

When calculating the increments in the structure function, we use pairs of points
separated by the periodic boundaries in $x$ and $y$. In the shearing box, the
horizontal periodicity conditions \citep[see][]{Hawley:1995} for a variable $f$ are 

\begin{equation}
\begin{aligned}
f(x,y,z) &= f(x + L_x, y - \delta y(t), z) \quad	&&(\text{boundary in }x), \\
f(x,y,z) &= f(x, y + L_y,z) 						&&(\text{boundary in }y), \\
\end{aligned}
\label{eq:bcs}
\end{equation}

where $\delta y(t) = \mathrm{mod}[q \Omega L_{x} t, L_{y}]$ is the time-varying 
offset between the shearing boundaries in $x$ (mapped to the range 
$0 \le \delta y < L_{y}$).
In order to conveniently include pairs of points located on different sides of the 
periodic boundary in $x$, we extend the computational domain in the $x$-direction by
its copy and shift it by $\delta y(t)$ to remove the discontinuity between the two
domains, as shown in Figure~\ref{fig:Figure1}.

\section[]{Spatial correlations}
\label{sect:spatial}
\begin{figure}
\centering
\includegraphics[width=0.47\textwidth]{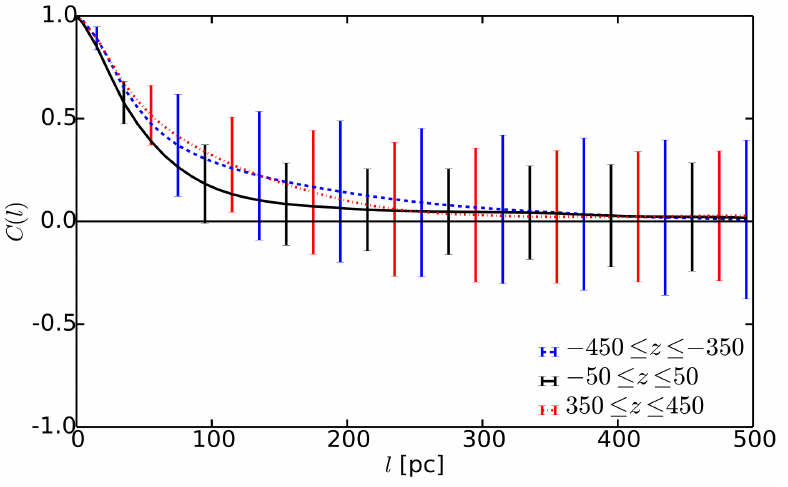}
\caption{
As in Fig.~\ref{fig:Figure2}, but for the
random speed in the warm gas.
\label{fig:Figure3}
}
\end{figure}

\begin{figure}
\centering
\includegraphics[width=0.47\textwidth]{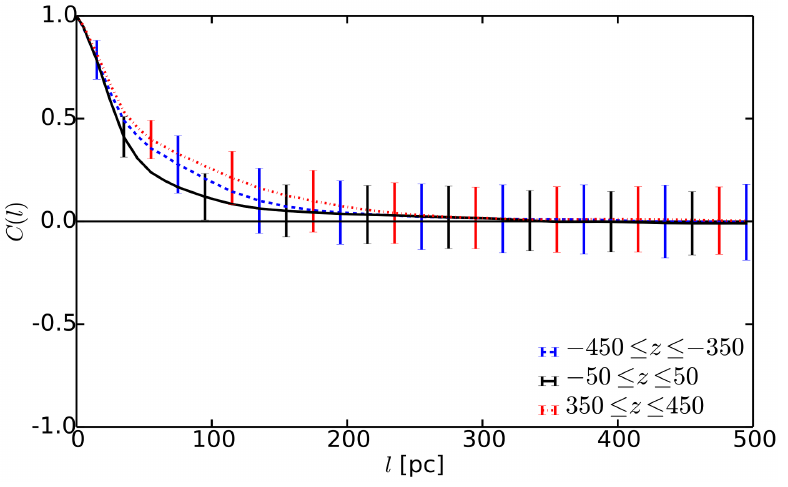}
\caption{As in Fig.~\ref{fig:Figure2}, but for the modulus of the 
random magnetic field in the warm gas. 
\label{fig:Figure4}
}
\end{figure}

As described above, we calculate the spatial structure- and correlation-functions 
for the random magnetic and velocity fields and the fluctuations in the gas number 
density separately for the warm and hot gas. The correlation functions are then used
to estimate the correlation lengths of these variables. Spatial correlations of the
Faraday depth and synchrotron emissivity are discussed in 
\S{\ref{sect:observables}}.

The results are shown in Table~\ref{tab:Table1} and Figure~\ref{fig:Figure2} for the
density fluctuations, Figure~\ref{fig:Figure3} for the random speed and
Figure~\ref{fig:Figure4} for the magnitude of the random magnetic field. The 
structure functions used to obtain the autocorrelation functions are only shown in
Figure~\ref{fig:Figure2}a: those for the other variables have a similar form. 
The magnitudes of the fluctuations in the variables and their correlations lengths 
are discussed in the next two sections.

The uncertainties of the root-mean-square (rms) values of various variables and 
their correlations lengths given in Tables~\ref{tab:Table1}, \ref{tab:Table2} and
\ref{tab:Table5} have been obtained as 95\% confidence intervals from weighted 
least-squares fitting of equation~\eqref{eq:l_fit2}, or for the gas density
equation~\eqref{eq:cf_fit}.
The weights used are the uncertainties of the values of the correlation function
rather than the standard deviations shown in the figures.

The uncertainties in the rms values and correlation lengths thus obtained are
underestimates of the true uncertainty as they do not take into account any
systematics errors, such as those arising from the uncertain value of the computed
structure functions at $l\to\infty$.

\subsection{Magnitude of the fluctuations}
\label{sect:n_spatial}

The rms magnitudes of the fluctuations are shown in Table~\ref{tab:Table1}, together
with the rms values of the relative fluctuations, 
$\langle (f'/\langle f\rangle_l)^2\rangle^{1/2}$ for a variable $f$; we stress that
the mean value $\langle f\rangle_l$ is a function of position. In the case of
velocity fluctuations, the average velocity, $\langle \vect{u}\rangle_l=0$, refers 
to the sheared frame, that is, includes the systematic outflow velocity, but not the overall rotation or the shear due to the galactic differential rotation.

In each phase, the standard deviation of the density fluctuations decreases with 
$|z|$ together with the average density. The relative magnitude of the fluctuations
also decreases, but more slowly.

As shown in Fig.~\ref{fig:Figure2}, density fluctuations are weakly 
anti-correlated in the range of scales $80 \leq l \leq 250 \p$ at each height, 
with the modulus of negativity for $C(l)$ significantly exceeding its uncertainty
(about $0.002$).
Therefore, the rms value and correlation length of the density fluctuations has been
obtained by fitting the form in equation~\eqref{eq:cf_fit} to the structure 
function. The parameters used in the cosine function were $k(l)=0.07l + 0.28$ at
$z=0\kpc$ and $k(l)=0.075l + 0.4$ at $|z|=0.4\kpc$.

A possible cause of such anticorrelation may be random shock waves propagating 
through the ISM. Then the density fluctuations can be expected to be correlated 
within distances comparable to the shock thickness (about $5\Delta x=20\p$ in the
simulations), whereas the anticorrelation arises from the systematic rarefaction
associated with a shock front. 
Another effect that may contribute to such anticorrelation is the presence of 
quasi-spherical supernova remnants 
(as are clearly visible in Fig.~\ref{fig:Figure1}), with gas density systematically 
lower than average within and around the bubbles and higher than average in their
shells.

\label{sect:u_spatial}
The rms random speed decreases with $|z|$ between $z=0$ and $|z|=400\p$. 
This is understandable since the Type II supernovae,
that drive most of the random flow, have a scale height of only $90\p$.
At larger heights, the rms $u'$ is $5\pm1\kms$ in the warm phase and 
$11\pm7\kms$ in the hot gas at $|z|=0.8\kpc$.

The magnitude of $\sigma_b$ in the simulations is below $\simeq5\mkG$ observed near
the Sun or in external galaxies \cite[][and references therein]{Beck:2016}. There
could be several reasons for this, including the relatively low magnetic Reynolds 
numbers in the simulations reducing fluctuation dynamo efficiency, or an 
underestimated averaging scale $\ell$. However, it is evident from Figure 6 of
\citet{Gent:2013b}, that its underestimation would not explain this.
Applying horizontal averaging, which is analogous to extending $\ell$ to 1\,kpc,
yields an increase of only 50\% in the saturated magnetic energy of the fluctuation
field.

\subsection{Correlation scales}
\label{sect:spatial_scales}
The correlation length of the density fluctuations in the warm gas shown in
Table~\ref{tab:Table1} decreases with $z$ in the range $|z| \leq 400 \p$, 
in contrast to the correlation lengths of the velocity and magnetic fields. 

In the simulations used here, shock-capturing diffusivities smooth shock fronts over
five mesh points, i.e., $20\p$. This shock-capturing smoothing may affect the
correlation lengths obtained, even though they are normally significantly larger 
than $20\p$. It may particularly affect the correlation length for the density
fluctuations at $|z|=400\p$, which is only $37\p$. 

The correlation length of the random velocity at the same height is significantly
larger. The corresponding correlation length of the random magnetic field is
intermediate between the two.

From the double rotation rate simulation, the results obtained for the correlation lengths and rms values are very similar to those in Table~\ref{tab:Table1}. 

\subsection{Taylor microscale}
\label{sect:microscale}
\begin{table}
\centering
\caption{
Estimates of the Taylor microscale obtained from fitting equation~\eqref{eq:curve_fit}
to the autocorrelation function of the random speed at $z=0$, for decreasing bin widths in $l$.
\label{tab:Table2}
}
\begin{tabular}{lcccc}
\hline\hline
Bin width [pc]	&12					&10					&8				&6\\
\hline
$\lambda$ [pc]	&$46\pm10$	&$45\pm10$	&$48\pm8$	&$40\pm11$\\
\hline
\end{tabular}
\end{table}

\begin{figure}
\centering
\includegraphics[width=0.47\textwidth]{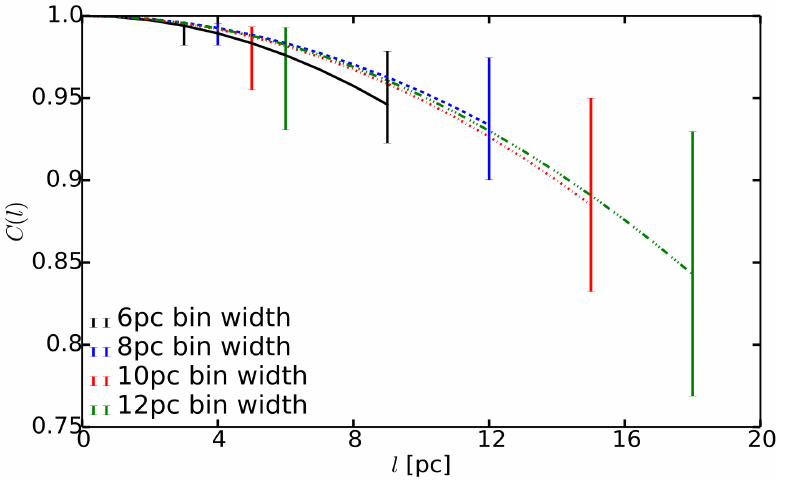}
\caption{Parabolic fits to the autocorrelation function for random speed
         $u^\prime$, averaged about the mid-plane; for bin widths 
         $6 \p$ (black, solid), $8 \p$ (blue, dashed), 
         $10 \p$ (red, dash-dotted), and 
         $12 \p$ (green, dash-triple-dotted).
         The autocorrelation function for each bin width is point plotted,
         with only the error bars shown.
}
\label{fig:Figure5}
\end{figure}

\begin{figure}
\centering
\includegraphics[width=0.47\textwidth]{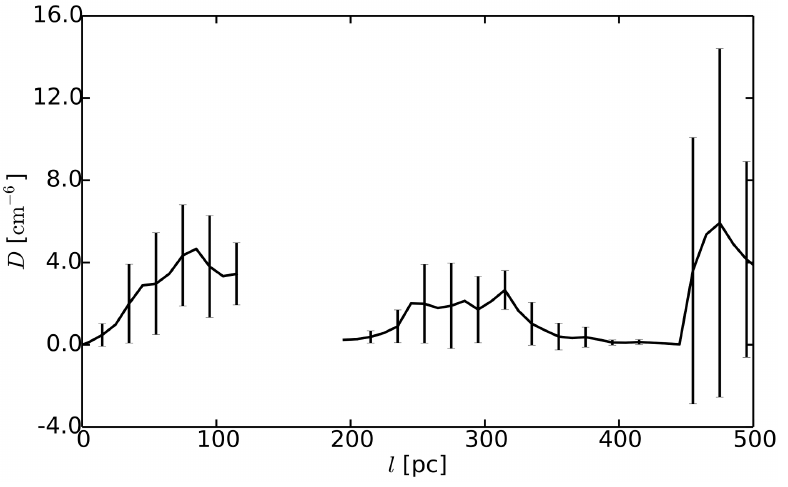}
\caption{Structure function for random density in the cold phase for $|z|\le50\p$
(the cold gas is practically absent at larger values of $|z|$).}
\label{fig:Figure6}
\end{figure}

\begin{table*}
\centering
\caption{
The root-mean-square (rms) values of the fluctuations, the relative rms values and correlation lengths of the fluctuations in gas density, speed, and magnetic field, where no phase separation has been applied, at the mid-plane $z=0$ and at $|z|=400\p$.
\label{tab:Table3}}
\begin{tabular}{@{}l cccc @{}c cccc  @{}c cccc@{}}%
\hline\hline
$|z|$         &\multicolumn{4}{c}{Density fluctuations}                  & &\multicolumn{4}{c}{Random speed}                   & &\multicolumn{4}{c}{Random magnetic field}               \\
               \cline{2-5}                                                  \cline{7-10}                                          \cline{12-15}                                           \\[-8pt]
              &rms               &rms           &$l_0$    &$\tilde{l}_0$ & &rms           &rms        &$l_0$    &$\tilde{l}_0$ & &rms             &rms           &$l_0$    &$\tilde{l}_0$ \\
$[\text{pc}]$ &[cm$^{-3}$]       &relative      &[pc]     &[pc]          & &[km/s]        &relative   &[pc]     &[pc]          & &$[\mu$G]        &relative      &[pc]     &[pc]          \\
\hline
0             &$0.305\pm0.001$   &$0.63\pm0.09$ &$44\pm2$ &$29\pm1$      & &$12.89\pm0.03$ &$1.2\pm0.5$ &$74\pm2$ &$63\pm1$     & &$0.582\pm0.001$ &$0.59\pm1.0$ &$51\pm1$ &$44\pm1$     \\
400           &$0.0604\pm0.0001$   &$0.33\pm0.03$ &$37\pm2$ &$27\pm2$  & &$3.65\pm0.01$ &$0.5\pm0.1$  &$117\pm3$ &$112\pm2$     & &$0.484\pm0.001$ &$0.39\pm0.05$ &$66\pm1$ &$58\pm1$      \\
\hline
\end{tabular}
\end{table*}

\begin{figure*}
\centering
\includegraphics[width=0.31\textwidth]{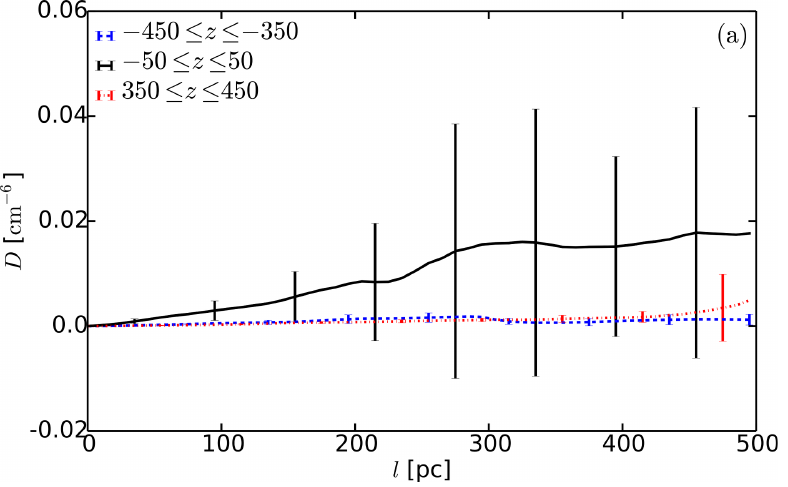}
\hfill
\includegraphics[width=0.31\textwidth]{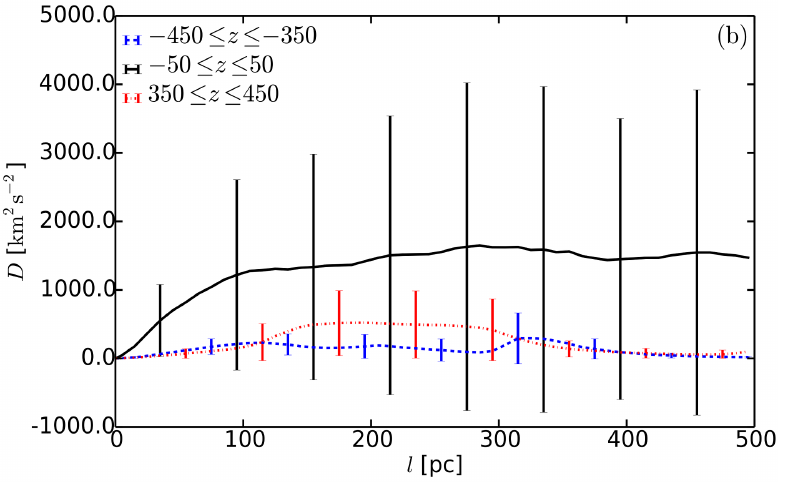}
\hfill
\includegraphics[width=0.31\textwidth]{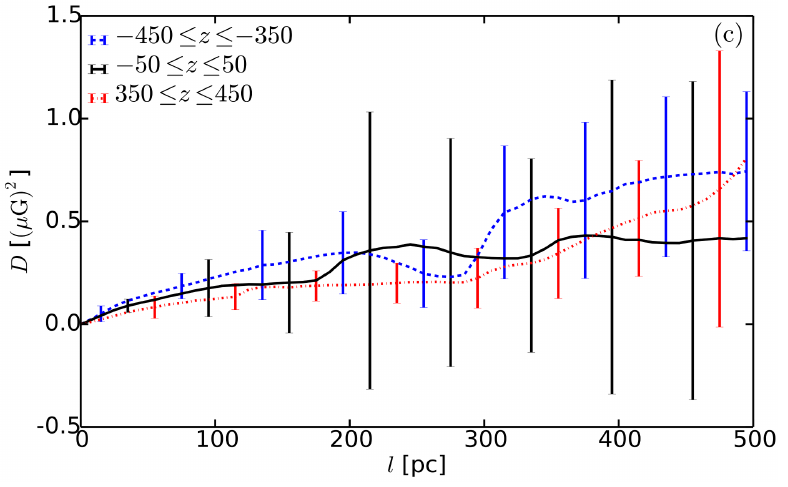}
\caption{Structure functions for (a) density fluctuations, (b) random speed and (c) random magnetic field strength 
in the hot gas, averaged about $z = -400\p$ (blue, dashed), $z = 0\p$ (black, solid), and $z = 400\p$ (red, dash-dotted).
\label{fig:Figure7}
}
\end{figure*}

\begin{figure*}
\centering
\includegraphics[width=0.47\textwidth]{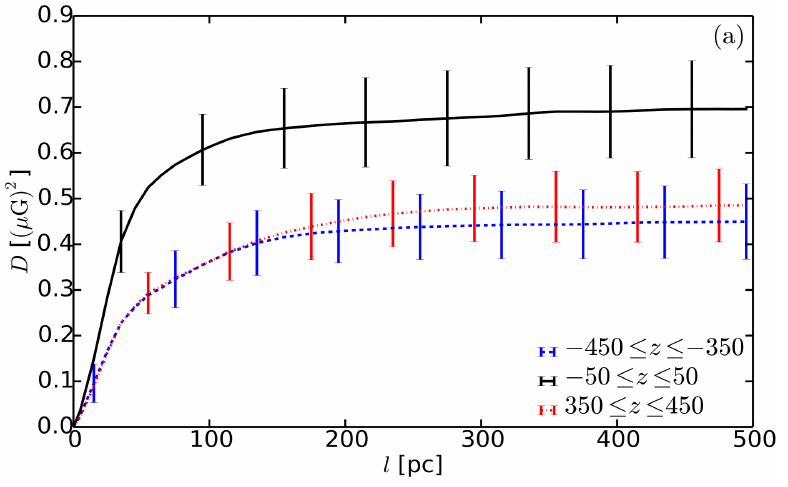}
\hfill
\includegraphics[width=0.47\textwidth]{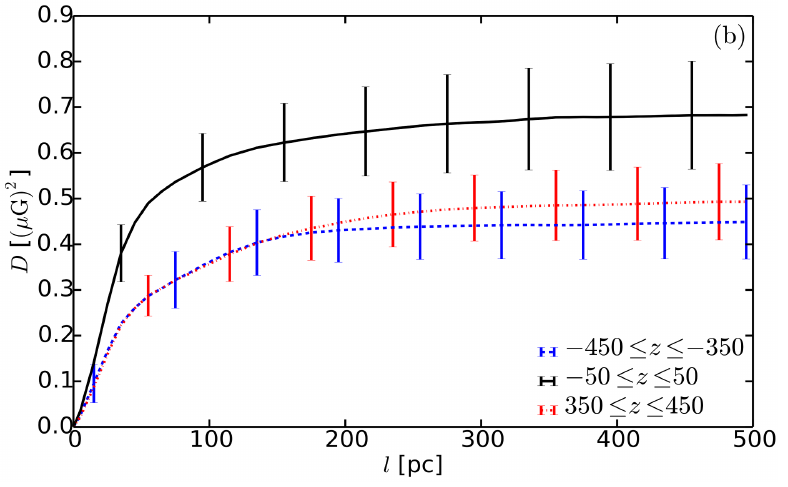}
\caption{Structure functions for random magnetic field strength in: 
(a)~the warm phase and (b)~the whole ISM.
\label{fig:Figure8}
}
\end{figure*}

The Taylor microscale, $\lambda$, characterizes the behavior of the correlation
function at small scales, $l\to 0$, and can be obtained by fitting the correlation
function near the origin to the form
\begin{equation}
C(l) \simeq 1 - \left(l/\lambda\right)^{2},
\label{eq:curve_fit}
\end{equation}
\citep[\S6.4 in][]{Tennekes:1972}.
The associated equality $\dd C/\dd l=0$ at $l=0$ holds for the correlation functions
of smooth (differentiable) random fields \citep{Monin:1975}. In numerical 
simulations, where the solutions at the smallest scales are controlled by the finite
numerical resolution $\Delta x$, one expects $\Delta x < \lambda < l_{0}$
\citep{Davidson:2004}.
The Taylor microscale of the random speed can be used to estimate the effective Reynolds number, $\mathrm{Re}$, in the simulations  \citep[e.g., \S3.2 in][]{Tennekes:1972},
\begin{equation}
\lambda \simeq 3l_{0} \mathrm{Re}^{-1/2}.
\label{eq:reynolds}
\end{equation}
Such an estimate includes all dissipation effects in an averaged manner, which can 
be difficult to estimate otherwise because of the extreme inhomogeneity of the
simulated ISM and numerical transport coefficients.
	
    Thus obtained, the Reynolds number is based on the correlation scale of the 
	random flow; the corresponding value based on the domain size 
	($1\kpc$), often quoted in the literature, is about $400$.

We fit equation~(\ref{eq:curve_fit}) to the correlation function $C(l)$ of the 
random gas speed at the three smallest values of $l$, including $C(l)=1$ at $l=0$, 
for bin width in $l$ of $6$, $8$, $10$, and $12 \p$. Figure~\ref{fig:Figure5} shows
the correlation functions obtained at $|z|\leq50\p$ and the fits.

The resulting estimates of $\lambda$, shown in Table~\ref{tab:Table2}, satisfy the
inequalities $\Delta x < \lambda < \lo$, providing us some confidence in the 
estimates of the correlation lengths discussed above. For $\lo = 60 \p$
(Table~\ref{tab:Table1}) and $\lambda=40\p$, we obtain an estimate of the effective
 Reynolds number in the simulations of order 20.
We also obtained similar results in a model with doubled velocity shear.

    The relatively low value of the effective Reynolds number is likely to be a
	consequence of the shock capturing numerical scheme used in the simulations, 
	where shock fronts are diffused over several grid points to be fully resolved. 
	The maximum Reynolds number achievable with the numerical resolution
	$\Delta x$ is of order 
	$\mathrm{Re}_\mathrm{max}\simeq(\lo/\Delta x)
	^{4/3}$ (assuming a power-law turbulent spectrum with a 
	slope of $5/3$).
    With $\lo=50\p$, this yields
	$\mathrm{Re}_\mathrm{max}\simeq30$, so the effective value of
	$\mathrm{Re}$ measured directly is not much smaller than the nominal value. 
	
    We also note that the value of the effective Reynolds number is likely to 
	be much lower than local values in diffuse gas because it includes strong
	numerical dissipation in shocks.

	A comprehensive analysis of vortex generation in the ISM by \citet{KGMS17} 
    identifies baroclinicity to be significantly the most efficient source of
    vorticity in SN driven turbulence.
    Vorticity is critical to dynamo action, and this conversion of potential into
    rotational flow may partly explain the persistence of the dynamo even at the
    relatively low Reynolds numbers, as compared to simulations which model SNe
    without thermal energy or viscous heating.

    The correlation scale of the random flow is
	controlled by the energy injection mechanism rather than the Reynolds number, so
	we believe that the modest value of the Reynolds number that our work shares with
	other comparable simulations does not affect our conclusions.

\subsection{Overall statistics and the cold and hot phases}
\label{sect:phases}
The results presented above are for the warm gas. The data for the cold gas at 
offsets beyond  
$l\simeq10\text{--}100\p$, 
the typical scale of the cold gas clouds, are scarce because the cold gas occupies a small fraction of the volume.
Furthermore, the numerical resolution of $4\p$ in our 
simulations restricts the quality of the modelling of the cold phase, localized 
in regions of order $10\p$ in size.
Additionally, our results only consider cold gas structures that are typical
of diffuse clouds, since we do not model the molecular gas 
(see Section~\ref{sect:parameters}).
 
Figure~\ref{fig:Figure6} only shows the cold phase results for the mid-plane, since
the cold gas is concentrated there, and results outside this region cannot be
statistically meaningful \citep[see][]{Gent:2012,Gent:2013a}. 
The structure functions for the hot phase fluctuate wildly and have large error bars
(see Figure~\ref{fig:Figure7}). This happens because the hot phase is extremely
variable within the relatively small computational box that we have.

A separate analysis for each ISM phase, feasible with simulated data, may not be
possible in observations. Therefore, we briefly discuss the statistical properties 
of the simulated ISM without separation by phase. The results are shown in
Table~\ref{tab:Table3}. 

As shown in Figure~\ref{fig:Figure8}, the structure and correlation functions of
magnetic fluctuations, $b$, for the whole ISM are almost identical to those in the
warm phase. This is also true of the gas density fluctuations $n'$. This similarity 
is reflected in the values of $l_b$, $l_{n'}$, $\sigma_b$, and $\sigma_{n'}$.
This is, of course, largely due to the large fractional volume of the warm phase. 
It is worth noting, however, that the density and magnetic field strength in the hot
phase are both lower than in the warm phase.

However, the values of $\sigma_{u'}$ and $l_{u'}$ for the whole ISM are 
significantly higher than in the warm phase. The larger values of $\sigma_{u'}$ for
the whole ISM can be attributed to the contribution of the hot gas that has higher
speed of sound and, correspondingly, higher random velocities.

\begin{figure}
\centering
\includegraphics[width=0.47\textwidth]{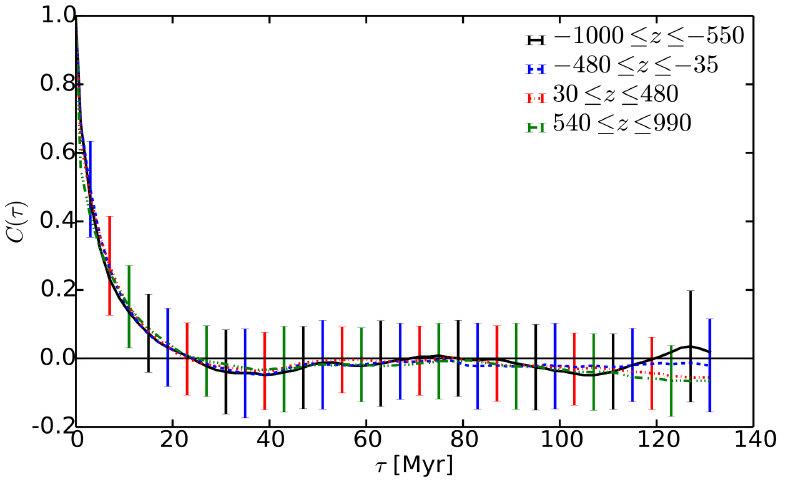}
\caption{Time autocorrelation functions, $C(\tau)$, for 
the random speed, for $-1000 \leq z \leq -550 \p$ (black, solid), $-480 \leq z \leq -35 \p$ (blue, dashed), 
$30 \leq z \leq 480 \p$ (red, dash-dotted), and $540 \leq z \leq 990 \p$ (green, dash-triple-dotted). 
The error bars denote the standard deviation of the individual 
contributions to the correlation function around the mean value.
For clarity, only every 
fourth error bar is shown at each curve.}
\label{fig:Figure9}
\end{figure}

\section{Time correlation}
\label{sect:time_correlation}
Unlike the correlation lengths of various observable quantities in the ISM, their
correlation \textit{times} cannot be obtained from observations. Because of this, 
the eddy turnover time $\tau=l_0/u_0$ is universally applied to interstellar
turbulence. 
However, the dynamics of interstellar turbulence involves a range of physical
processes having distinct time scales, which may make the eddy turnover time
inappropriate as an estimate of the correlation time. 
Nonlinear Alfv\'en wave interactions, shock-wave turbulence and fluctuation dynamo
action, among other phenomena, are likely to affect the correlation time and make it
different for different variables.

Similarly to correlation lengths, the correlation times can be different in the warm
and hot phases. However, this difference is harder to capture since each parcel of
warm or hot gas moves around. Therefore, we can only obtain correlation times 
averaged over the ISM phases.

We consider arguably the most important of the time correlations, that of the random
velocity. For this purpose, we use time series of the magnitude of the random 
velocity measured at an array of fixed points in $32$ planes in $z$, separated by 
$64 \p$; within each plane, there are $64$ positions separated by $100 \p$ in 
$x$ or $y$.

From this data, we can calculate the temporal structure function, and then
autocorrelation function $C(\tau)$, from which we  obtain the correlation time 
$\tau_0$,
\begin{equation}
\tau_{0} = \int_0^\infty C(\tau) \; \mathrm{d}\tau \;.
\label{eq:time}
\end{equation}
We fit the form in equation~\eqref{eq:l_fit1} to $C(\tau)$ to estimate $\tau_0$.

The autocorrelation functions are shown in Figure~\ref{fig:Figure9} for four 
distances from the mid-plane, and the correlation times can be found in
Table~\ref{tab:Table4}: $\tau_0 \approx 5 \Myr$ with little variation with $|z|$. 
Since the fractional volumes of the warm and hot gas vary significantly with $|z|$,
this suggests that both phases have similar correlation times.

With the velocity correlation length and speed in the warm gas at $z=0$ of $60\p$ 
and $8\kms$, respectively (from Table~\ref{tab:Table1}) the kinematic time scale 
(`eddy turnover time') is of order 
$\tau_\mathrm{eddy}=l_{u'}/\sigma_{u'}\simeq 8\Myr.$
At $|z|=400\p$, we similarly have $\tau_\mathrm{eddy}\simeq30\Myr$ in the warm gas.

According to the model of interstellar shock-wave turbulence of \citet{Bykov:1987}, 
the separation of primary shock fronts driven by supernova explosions depends on 
their Mach number $M$ as

\begin{table}
\centering
\caption{
The correlation time of the random speed at various heights in the simulation domain. \label{tab:Table4}}
\begin{tabular}{c c}
\hline\hline
$z$ & $\tau_0$ \\
$[\text{pc}]$ & [Myr] \\
\hline
$-1.000 \leq z \leq -0.550$ & $4.6 \pm 0.6$ \\
$-0.480 \leq z \leq -0.035$ & $4.9 \pm 0.5$ \\
$0.030 \leq z \leq 0.480$ & $4.9 \pm 0.7$ \\
$0.540 \leq z \leq 0.990$ & $4.6 \pm 1.0$\\
\hline
\end{tabular}
\end{table}

%
\begin{equation}
L_\text{shock} \simeq 4 M^{4.5}\p\,,
\label{eq:shocksep}
\end{equation}
where the galactic supernova rate of $0.02\yr^{-1}$ has been adopted. The primary
shocks dominate over weaker secondary shocks for $M\gtrsim1.2$, which leads to
$L_\text{shock}\simeq10\p$. The corresponding time between crossings of a given
position by shock fronts, which is expected to destroy time correlations, then 
follows as $\tau_{\rm{shock}}=L_{\rm{shock}}/c\simeq0.7\Myr$, where $c=14\kms$ is 
the magnetosonic speed in the warm gas (assuming equality of the sound and Alfv\'en
speeds).

In the simulations with double rotation rate, the velocity correlation rate and 
speed at the mid-plane in the warm phase change to $58 \p$ and $8 \kms$, resulting 
in the eddy turnover time of $\tau_{\rm eddy} \approx 7 \Myr$, whereas 
$\tau_{\rm shock}$ remains unchanged.

Since the estimate of $\tau_0$ that we have does not distinguish between the hot 
and warm phases, it depends on both the kinematic and shock-crossing time scales 
in each phase (and also the Alfv\'en time scale, but this is close to the kinematic
time scale since the magnetic and kinetic energy densities are comparable).
All these time scales are of the same order of magnitude, so more careful estimates 
of the correlation time are required to clarify the physical nature of the time
correlations in the simulated ISM.

It is plausible that the correlation time reflects both time scales and 
$\tau_0^{-1}\simeq \epsilon
\tau_{\rm eddy}^{-1} + (1-\epsilon)\tau_\text{shock}^{-1}$ 
with a certain constant $\epsilon$. With
$\tau_0=5\Myr$, $\tau_\text{\rm eddy}=7\Myr$ and $\tau_\text{shock}=1\Myr$, 
we obtain $\epsilon \simeq 0.9$, so the shock
waves contribute about 10\% to the random flow in this sense.

\begin{figure*}
\centering
\includegraphics[width=0.32\textwidth]{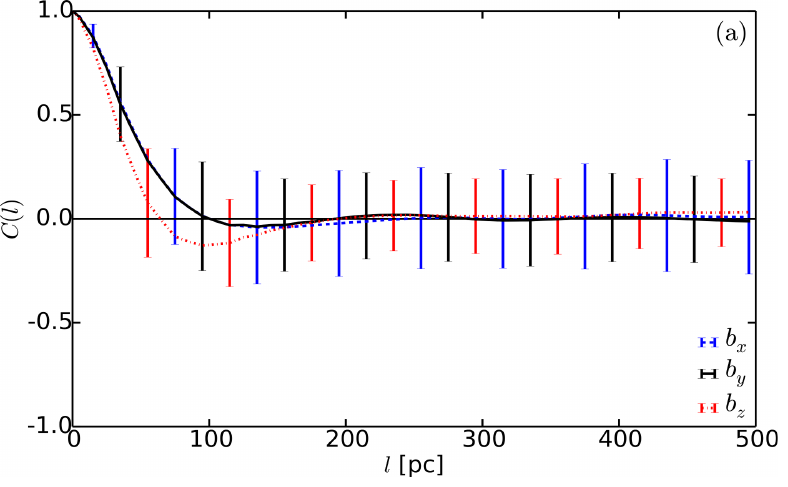}
\hfill
\includegraphics[width=0.32\textwidth]{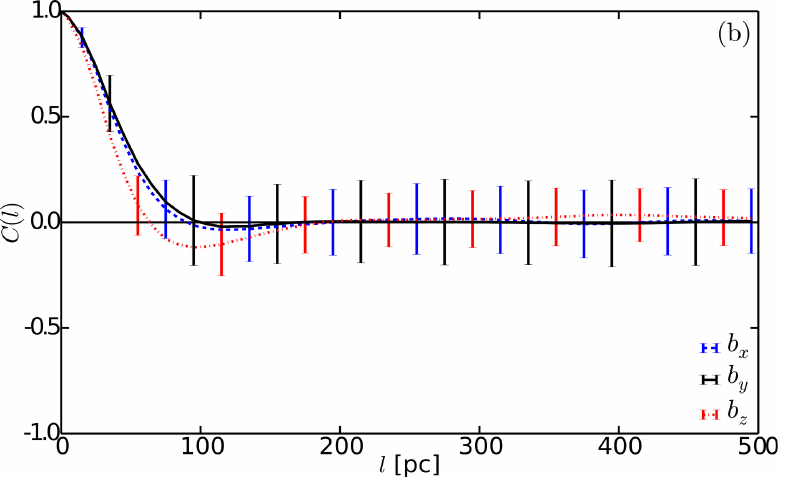}
\hfill
\includegraphics[width=0.32\textwidth]{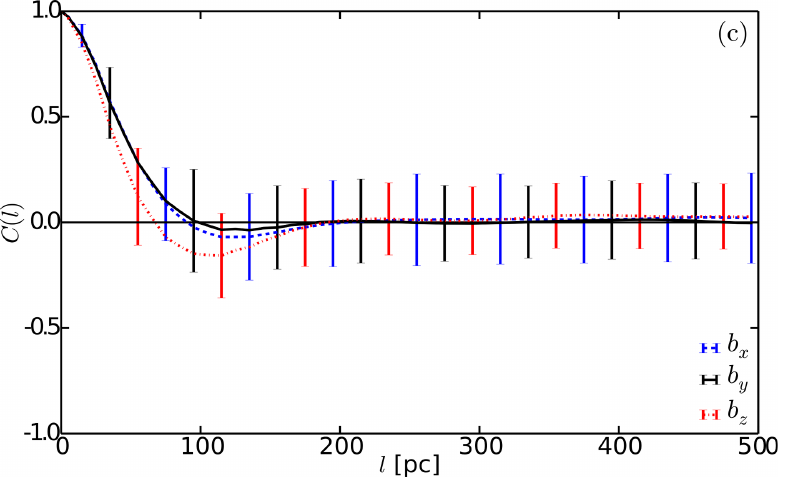}
\caption{Autocorrelation functions, $C(l)$ for magnetic field components, averaged at heights; a) $z = -400 \p$, b) $z = 0 \p$ and c) $z = 400 \p$.
\label{fig:Figure10}
}
\end{figure*}

    The time autocorrelation function of Fig.~\ref{fig:Figure9} appears to
	vary around the zero level at a time scale of about $70\Myr$.
	Although the accuracy of the autocorrelation values is higher 
    than suggested by the scatter of the data points around the mean 
    values shown by the error bars, the statistical significance of these 
    variations is unclear. Physical interpretation  of the time correlation 
    function is also hampered by the fact that we do not have time series for 
    the variables in the warm and hot phases separately.
    
    We note, however, that the apparent time scale of the variations is close 
	to the period 
	$(2\pi \lambda/g_{z})^{1/2}$ 
	of gravity waves of wavelength 
	$\lambda=1\kpc$ in the Galactic gravity field,
	$g_{z}\simeq4\times10^{-9}\cm\s^{-2}$.
	Oscillatory large-scale horizontal vortical flows have been found by 
	\citet{KGMS17} in similar simulations without a magnetic field. 
	For the parameters relevant to the current study, the period of these 
	oscillations is about $50\Myr$. 
	Extending this analysis over a range of 
	rotation rates, SN rates and forms of the gravitational potential may clarify 
	the significance of the pattern in the temporal correlation function apparent 
	in Fig.~\ref{fig:Figure9}.

\begin{table}
\centering
\caption{The rms values for each component of the random magnetic field, 
denoted $b_{0i}$ for $i\in (x,y,z)$, and their correlation lengths.
\label{tab:Table5}
}
\begin{tabular}{l cc c cc c cc}
\hline\hline
	   &\multicolumn{2}{c}{$b_{0i}$ [nG]} &&\multicolumn{2}{c}{$l_0 \, [\rm pc]$} &&\multicolumn{2}{c}{$\tilde{l}_0 \, [\rm pc]$}\\
	   \cline{2-3}																		\cline{5-6}																			\cline{8-9}\\[-6pt]
$|z|$ [pc] 	&$0$ 						&$400$  								&&$0$			&$400$					                       &&$0$ 						&$400$\\
\hline
$b_x$ 	   	&$549\pm1$	 & $432\pm1$   				&&$46\pm2$			&$51\pm3$																	&&$33\pm1$ 	&$32\pm1$\\
$b_y$	   		&$676\pm1$	 & $524\pm1$ 				                        	&&$41\pm1$	&$47\pm3$ 	  	&&$36\pm1$ 	&$36\pm1$\\
$b_z$	   		&$692\pm1$	 & $667\pm2$ 	  					                      &&$59\pm3$	&$65\pm3$		&&$16\pm1$ 	&$14\pm1$\\
\hline
\end{tabular}
\tablecomments{The correlation lengths $l_0$, using Eq.\,\eqref{eq:cf_fit}, and $\tilde{l}_0$ are calculated as in Table~\ref{tab:Table1}.}
\end{table}

\section{Anisotropy of the magnetic field}
\label{sect:aniso}
In the analysis above, we neglected any anisotropy of the random magnetic field in 
the horizontal planes. This is justifiable since, at the scales of interest 
(from a few parsecs to about $100\p$), the expected anisotropy is only moderate 
(see below). 
However, the anisotropy of magnetic fields is of high physical significance 
as it reflects the dynamics of MHD turbulence 
with and without a global mean magnetic field
(\citet[][]{Goldreich:1997,Brandenburg:2013} and references therein; 
see also \citet[][]{Cho:2000,Cho:2002a,Cho:2003b,Mallet:2016} and \citet{Oughton:2016}).
It also reflects the effects of galactic differential rotation 
and compression of the random magnetic field in shocks.
The anisotropy of interstellar magnetic fields can contribute significantly 
to the polarized radio emission of galaxies 
\cite[e.g.,][]{Sokoloff:1998,Beck:2016}.
In this section, using the structure and autocorrelation
functions, we discuss
individual components of the random magnetic field, $\boldsymbol{b}=(b_x, b_y,b_z)$
denoting their rms values $b_{0x}$, $b_{0y}$ and $b_{0z}$.

As shown in Table~\ref{tab:Table5}, the three components of $\boldsymbol{b}$ are
somewhat different in magnitude. The vertical, $z$-components is the largest at all
heights, whereas the radial ($x$) random field is the weakest.

All three components of magnetic field have negative autocorrelation 
near $l=100\p$, stronger for $b_z$ than for $b_x$ and $b_y$. 
This appears to be a consequence of the solenoidality of magnetic field: since 
magnetic lines must be closed, magnetic field must, on average, change its 
direction at a length scale comparable to its correlation length.

An enhanced  azimuthal ($y$) component is a result of the large-scale velocity shear
due to differential rotation that produces $b_y$ from the radial field $b_x$, so 
that $\partial b_y/\partial t\simeq q\Omega b_x$ and then
\citep[e.g.,][]{Stepanov:2014}
\begin{equation}
b_{0y} \simeq (1 + q \Omega \tau_0)b_{0x}.
\label{eq:aniso_by}
\end{equation}
For $q=+1$, $\Omega = 25 \kms \kpc^{-1}$ and
$\tau_0 = 5 \Myr$, this yields $b_{0y}/b_{0x} \simeq 1.2\text{--}1.3$,
in agreement with the estimates of Table~\ref{tab:Table5} at $z=0$.

\begin{figure}
	\centering
	\includegraphics[width=0.45\textwidth]{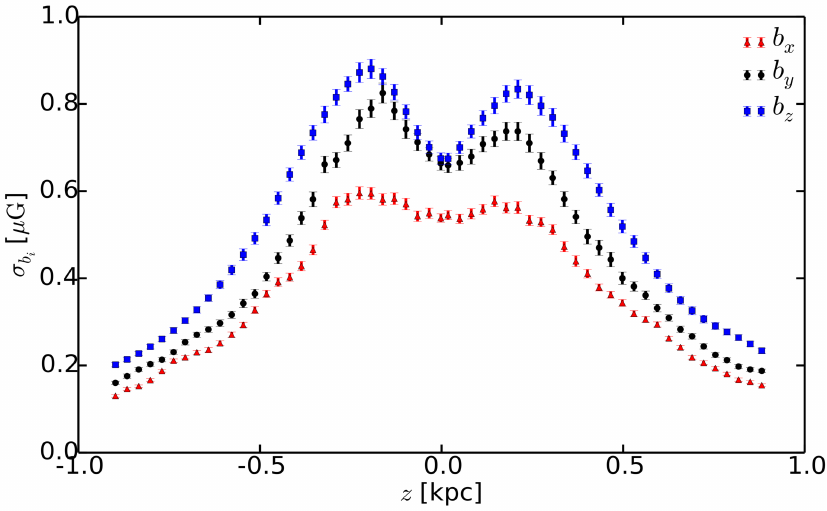}
	\caption{
		The rms values of the components of the random magnetic field vector as functions
		of distance to the mid-plane: the radial $b_{0x}$ (red, triangles), azimuthal $b_{0y}$ (black, circles),
		and vertical $b_{0z}$ (blue, squares) random magnetic fields.
	}
	\label{fig:Figure11}
\end{figure}

The vertical component of the magnetic field is similarly enhanced beyond isotropy
due to the stretching of the horizontal magnetic field by vertical velocity $u_z$ 
that varies at a scale $\lo$ and yet has a mean part $\overline{u}_z\simeq2\kms$ at 
$|z|\lesssim200\p$:
$\partial b_z/\partial t \simeq b_x\partial u_z/\partial x 
+ b_y\partial u_z/\partial y$.
Unlike the stretching of the radial magnetic field by the large-scale velocity shear,
this is a random process, so the rms vertical magnetic field grows as $t^{1/2}$. 
With the radial field $b_x$ representing the isotropic background, this leads to the
estimate
\[
\frac{b_{0z}}{b_{0x}}\simeq \left[1+\frac{\tau_0\overline{u}_z}{l_0}\left(1+\frac{b_{0x}^2}{b_{0y}^2}\right)\right]^{1/2}
\simeq1.2\,,
\]
in a reasonable agreement with the estimates of Table~\ref{tab:Table5}. Since the
vertical component of the random magnetic field is produced from both of its
horizontal components, the $z$-component is the strongest one.

An important radio astronomical consequence of the magnetic anisotropy is 
polarization of the synchrotron emission. If our simulation domain was observed from
the top or bottom (i.e., along the $z$ direction) the observed degree of 
polarization due to the random magnetic field alone would be
\citep{Laing:1981,Sokoloff:1998, Sokoloff:1999}
\[
\label{eq:anisoPI}
p = p_0\frac{|b_{0x}^2 - b_{0y}^2|}{b_{0x}^2 + b_{0y}^2}\approx 0.15,
\]
where $p_0\approx 0.7$ is the maximum intrinsic degree of polarization,
and we have neglected, for the sake of the argument, both depolarization effects and
the average magnetic field. Such a degree of polarization is comparable to that
observed in spiral galaxies, suggesting that the anisotropy of the interstellar 
random magnetic fields needs to be allowed for in the interpretations of radio
polarization observations of spiral galaxies \citep[cf.][]{Beck:2016}.

The correlation lengths of the magnetic field components are given in 
Table~\ref{tab:Table5} (for comparison with Table~\ref{tab:Table1}).
Because of the stretching of radial magnetic field by differential rotation that
produces a stronger azimuthal field, we might expect the azimuthal correlation 
length to be larger than the radial one \citep{Moffatt:1967,Terry:2000}, contrary to
the results in Table~\ref{tab:Table5}, where the correlation lengths for $b_{x}$ and $b_{y}$ are of similar magnitude. However, the correlation lengths were calculated
using isotropic horizontal position lags, whereas  azimuthal ($y$) and radial ($x$)
lags should be considered separately to detect the expected difference in the
correlation lengths in the two directions. Such a refined calculation requires a
larger data domain to provide sufficient statistics.
\citet{Houde:2013} find that $l_{0y}\approx 1.8 l_{0x}$ for the random magnetic 
field, i.e., the magnetic correlation length approximately along the mean-field
direction ($y$ in our case) is about twice that in the perpendicular direction, 
and this ratio is similar to the ratio of $b_{0y}/b_{0x}$ found by these authors 
from depolarization of the synchrotron emission. 
The vertical magnetic field component has significant anticorrelation at
$l\approx100\p$, shown in Figure~\ref{fig:Figure10}, which results in very 
different values of $l_0$ and $\tilde{l}_0$, similar to $n'$.

As shown in Figure~\ref{fig:Figure11}, individual components of the random 
magnetic field vary differently with $|z|$. As with the mean magnetic field,
the rms means first increase with distance from the mid-plane until 
$|z|\approx 200\p$, and only then decrease. As suggested above, both $b_y$ and $b_z$
are enhanced, in comparison with $b_x$, by the horizontal velocity shear and random
vertical flows, respectively; correspondingly, $b_{0y}$ and $b_{0z}$ increase with 
$|z|$ faster than $b_{0x}$ at $|z|\lesssim200\p$, but then decrease with 
$|z|$ following the decrease in $b_{0x}$.
At $|z| \geq 300 \p$, each component of $\boldsymbol{b}$ decreases nearly
exponentially with the scale height of about $450 \p$.

Simulations with double rotation rate produce similar results.

\section{Observable quantities}
\label{sect:observables}
The main observational tools employed in the analysis of interstellar MHD turbulence
are Faraday rotation and synchrotron emission, both total and polarized. Their
statistical properties and their relation to the underlying random distributions of
magnetic fields, gas density and cosmic rays have received significant attention,
both observationally and theoretically (see references in 
\S\ref{sect:introduction}). Here we discuss correlation properties of the observable
quantities in the simulated ISM. Given that magnetic field and gas density can have
different correlation functions, and can be correlated with each other
\citep{Beck:2003}, statistical properties of the observable quantities are difficult
to predict with confidence. 

Both Faraday rotation and synchrotron emission depend on the relative orientation of
the large-scale magnetic field and the line of sight. The mean magnetic field in the
simulations used here is predominantly horizontal and its $y$-component is the
strongest \citep{Gent:2013a,Gent:2013b}. Exploring the observational appearance of 
the simulated volume from various vantage points will be our goal elsewhere; here we
only discuss the properties of fluctuations in Faraday rotation and synchrotron
emission using just one direction of `observation'.

\begin{figure}
	\centering
	\includegraphics[width=0.45\textwidth]{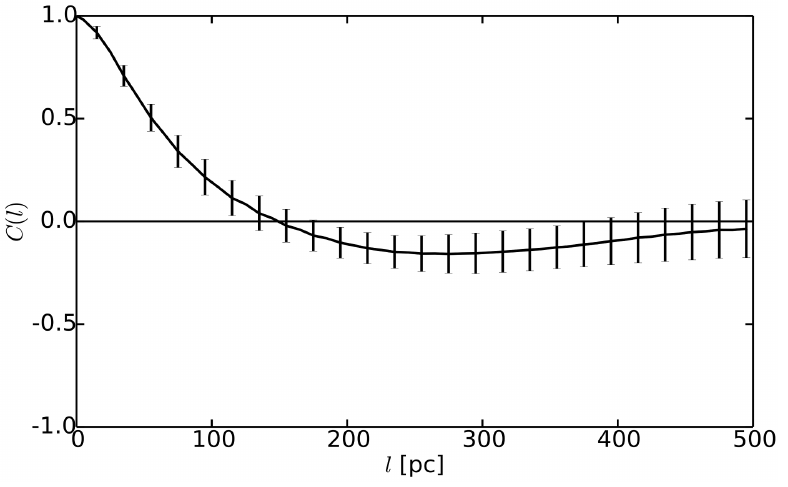}
	\caption{The autocorrelation function of the Faraday depth $\phi(x,y)$. 
		The error bars represent the scatter of the data points around the mean values shown with solid line. 
		\label{fig:Figure12}
	}
\end{figure}

\begin{table}
\centering
\caption{
The rms values and the
correlation lengths for electron density $n_{e}$.
\label{tab:Table6}
}
\begin{tabular}{l cc cc}
\hline\hline
    $|z|$ &&rms               &&$l_0$\\
    \\[-6pt]
    [pc]  && [$\rm cm^{-3}$]  &&$[\rm pc]$\\
\hline
$0$   &&$0.2560\pm0.0007$ &&$59\pm3$\\
$200$ &&$0.1665\pm0.0008$ &&$61\pm5$\\
$400$ &&$0.0530\pm0.0003$ &&$80\pm6$\\
$600$ &&$0.0208\pm0.0002$ &&$93\pm8$\\
$800$ &&$0.0083\pm0.0001$ &&$83\pm7$\\
\hline
\end{tabular}
\end{table}

\begin{figure*}
\centering
\includegraphics[width=0.47\textwidth]{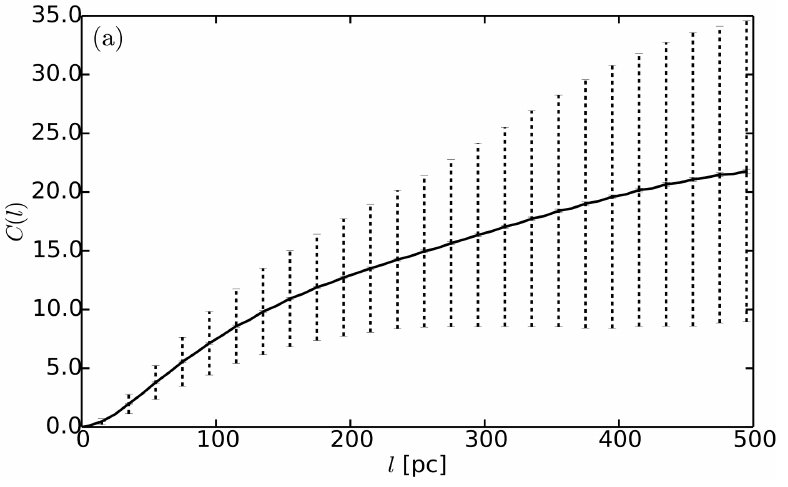}
\hfill
\includegraphics[width=0.47\textwidth]{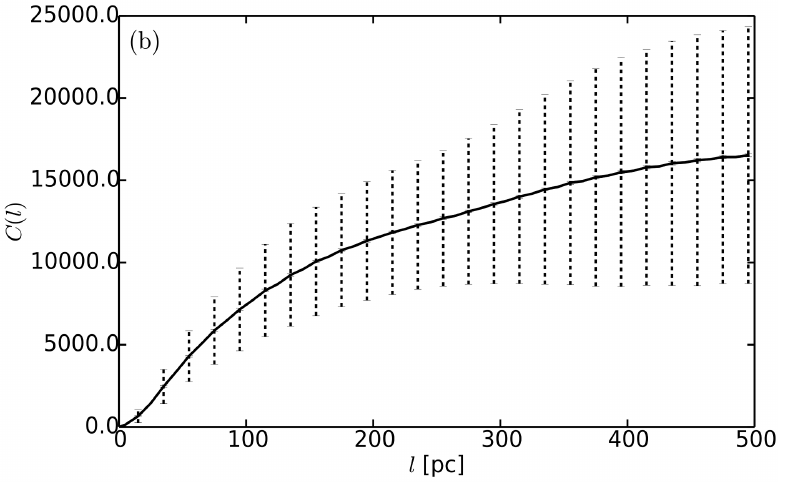}
\caption{
Structure functions of the synchrotron intensity assuming (a) constant cosmic ray density, 
$I$; and (b) local energy equipartition, $I_\mathrm{eq}$.
Vertical dashed lines represent
the standard deviation of the individual data points around the mean value shown with solid line,
whereas error bars show the accuracy of the mean. The computational volume is `observed' 
at roughly the right angle to the mean magnetic field, similarly to Milky Way observations at high 
Galactic latitudes.
\label{fig:Figure13}
}
\end{figure*}

\subsection{Faraday depth}
\label{sect:faraday}
The Faraday depth of a magneto-ionic region is an integral along the 
line of sight, assumed here to be along the $z$-direction for convenience:
\begin{equation}
\phi(x,y) = 0.81 \int_{-L_{z}}^{L_{z}} n_\text{e}  B_{z} \, \mathrm{d}z \; \radm,
\label{eq:faraday}
\end{equation}
where $n_\mathrm{e}$ is the number density of thermal electrons in cm$^{-3}$, 
$B_z$ is the line-of-sight component of magnetic field in $\mu$G, distance $z$ 
is in $\rm pc$, and $L_z$ is the half-size of the computational domain along $z$.
Since the mean magnetic field is nearly horizontal, the mean value
of $B_z$ is close to zero together with the mean Faraday depth 
along this direction.

Our simulations do not include gas ionization and only provide total gas density 
$n$. Since interstellar plasmas can be far from ionization equilibrium
\citep{deAvillez:2012a,deAvillez:2012b}, we obtain thermal electron density from
a heuristic relation that ensures that the mean electron number density is about
$0.03\cm^{-3}$ and the gas is fully ionized at $T\gtrsim10^{5}\K$:
\begin{equation}
n_\mathrm{e} = n \left[\frac{\mathrm{arctan}(T/10^3 \K - 10)}{\pi} + \frac{1}{2} \right].
\label{eq:electron_density}
\end{equation}
Since observations do not distinguish between different ISM phases, the Faraday 
depth has been computed for the whole computational domain.

The autocorrelation function of the Faraday depth is shown in
Figure~\ref{fig:Figure12}. 
Its correlation length, $l_{\phi}=122 \pm 12\p$ is significantly greater
than the correlation length of electron density, $60\p$ at the midplane 
increasing to $80\p$ at $|z|=800\p$ (Table~\ref{tab:Table6}),
and the vertical random magnetic field, $60\p$ (Table~\ref{tab:Table5}). 
We note that the \textit{mean} component of $B_z$ is negligible, so that the
mean value of the Faraday depth is close to zero, 
$\langle\phi\rangle=2.88\pm6.42\,\text{rad\,m}^{-2}$.

As discussed by \citet{Beck:2003}, the magnitude of Faraday rotation depends on the
correlation between magnetic field and thermal electron density. To clarify their
relation in our simulations, we computed the cross-correlation coefficient between
$n_\text{e}$ and $B_z$ separately for the warm and hot gas:
\begin{equation}
r=\frac{\overline{(n_\text{e}-\overline{n}_\text{e})\left(B_{z}-\overline{B}_{z}\right)}}{\overline{(n_\text{e}-\overline{n_\text{e}})^2}^{1/2}\, 
\overline{(B_z-\overline{B}_z)^2}^{1/2}}\,
\label{eq:cross_corr}
\end{equation}
where the overbar denotes an average taken over the volume occupied by the phase. 
The results, averaged over the snapshots, confidently suggest that the two variables 
are uncorrelated: $r=0.02\pm0.02$ in the warm gas and $0.07\pm0.04$ in the hot 
phase.
	
    The autocorrelation of $\phi$ is negative at 
    $l\gtrsim150\p$. Both magnetic field (Section~\ref{sect:aniso}) 
    and gas density have negative autocorrelation at these scales 
    (Fig.~\ref{fig:Figure2}). Quantitative assessment of this feature
	should await a more detailed analysis of the ionization structure of the 
	modelled ISM, but we note that this behaviour can have important implications 
	for the interpretation of radio polarization observations of the ISM, in terms 
	of parameters of interstellar turbulence.
	
\begin{figure*}
\centering
\includegraphics[width=0.47\textwidth]{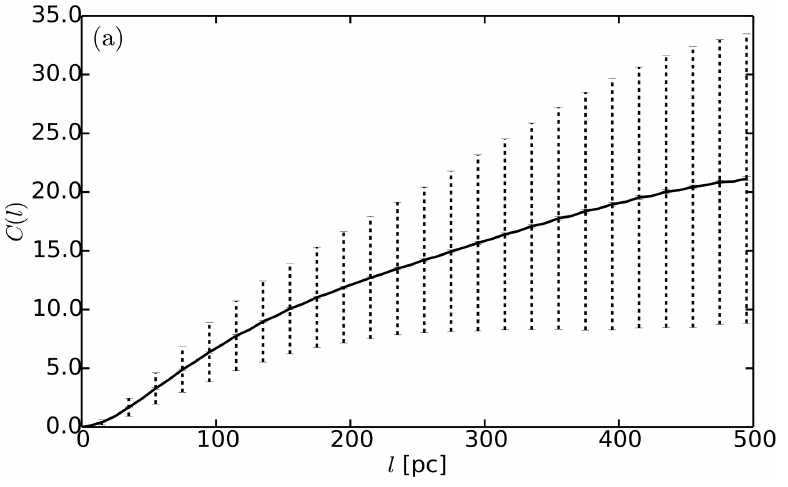}
\hfill
\includegraphics[width=0.47\textwidth]{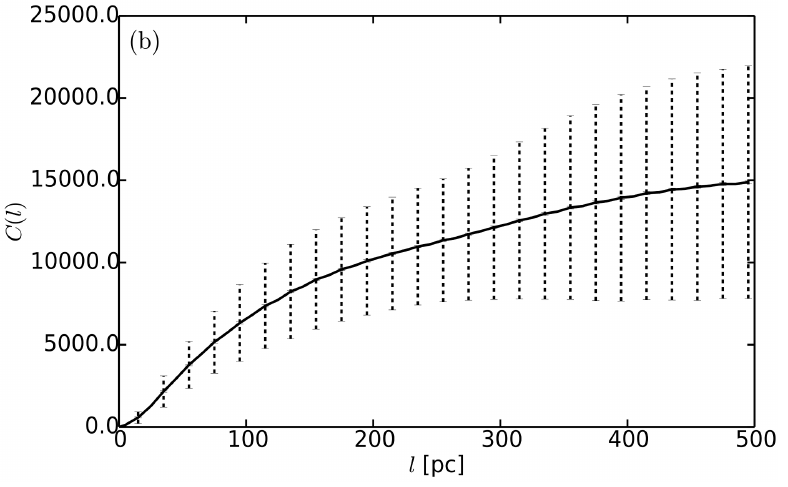}
\caption{As in Figure~\ref{fig:Figure13} but for polarized intensity assuming (a) constant cosmic 
ray density, $P$, and (b) local energy equipartition, $P_{\rm eq}$.
\label{fig:Figure14}
}
\end{figure*}

\subsection{Synchrotron intensity}
\label{sect:synchrotron}
Statistical properties of the synchrotron intensity are sensitive to the relation
between the distributions of cosmic ray electrons, $n_\text{cr}$, and magnetic 
field. Cosmic rays \citep{Berezinskii:1990} have a high diffusivity of order
$3\times10^{28}\cm^2\s^{-1}$, so their diffusion length over the confinement time 
of $10^6\yr$ is of order $1\kpc$. Thus, it can be expected that cosmic rays are 
distributed much more homogeneously than magnetic fields, but the assumption 
of a local energy equipartition (or pressure balance) between cosmic rays and 
magnetic fields is often used in interpretations of synchrotron observations
\citep[e.g.,][]{Beck:2005b}.
We note that analysis of synchrotron fluctuations in spiral galaxies 
suggest that cosmic ray electrons and magnetic fields can be slightly 
anti-correlated \citep{Stepanov:2014}. Fluctuations of synchrotron intensity can
provide information about interstellar turbulence 
\citep{Lazarian:2012,Lazarian:2016}.
Here we discuss the synchrotron intensity fluctuations implied by our ISM 
simulations. 

The synchrotron intensity, in arbitrary units, is obtained by integration along the
$z$-axis (so that the mean magnetic field is mostly perpendicular to the 
line of sight), 
\begin{equation}
I(x,y) = \int_{-L_z}^{L_z} n_\text{cr} \left(B_x^2+B_y^2\right) \, \mathrm{d}z,
\label{eq:synchrotron}
\end{equation}
using two alternative assumptions about cosmic ray distribution $n_\text{cr}$:
\[
n_\text{cr}	=\text{const}\,, 
\]
and
\[
n_\text{cr} \propto  B^2\,. 
\]
As with the Faraday depth, we do not consider other lines of sight through the computational domain.

The Stokes parameters, at wavelengths short enough that Faraday rotation is 
negligible, are similarly obtained as
\begin{align}
Q(x,y) &= \int_{-L_z}^{L_z}\cos(2\psi_0) n_\text{cr}(B_x^2+B_y^2) \, \mathrm{d}z \label{eq:stokes1}, \\
U(x,y) &= \int_{-L_z}^{L_z}\sin(2\psi_0) n_\text{cr}(B_x^2+B_y^2) \, \mathrm{d}z \label{eq:stokes2},
\end{align}
where $\psi_0 (\boldsymbol{x})$ is the intrinsic polarization angle perpendicular 
to the local magnetic field in the $(xy)$-plane, calculated as 
$\psi_0 = \pi/2+\arctan(B_y / B_x)$. The polarized intensity follows as
\begin{equation}
P(x,y) = \sqrt{Q^2 + U^2}.
\label{eq:polarised_intensity}
\end{equation}
The structure functions of the total and polarized synchrotron intensities under 
both assumptions about the cosmic ray distribution are shown in
Figures~\ref{fig:Figure13} and \ref{fig:Figure14}, respectively. They clearly have a
more complicated form than those of the magnitude of the random magnetic field shown
in Figure~\ref{fig:Figure4}. 
This is not surprising since the mean field is a function of position, and hence
contributes to the structure and correlation functions. In particular, the 
systematic increase of the structure function at large values of $l$ 
results from this contribution.  
The contribution from the mean field needs to be subtracted from the structure
function before any further analysis could be done. 
We postpone such analysis to simulations that include cosmic rays. 

A notable feature of the results illustrated in Figures~\ref{fig:Figure13} and
\ref{fig:Figure14} is the rapid increase in the scatter of the data points and the
deterioration of the accuracy of the structure function estimates as the lag $l$
becomes larger than about $200\p$. 
This is understandable since the synchrotron emissivity depends on relatively 
high power of the fluctuating magnetic field. Observations in the Milky Way can
be especially strongly affected,  because even within a narrow telescope beam
the divergence of the lines of sight can be as wide as hundreds of parsecs
at some distance from the Sun \citep{Cho:2002b,Cho:2003a,Cho:2010}.

In the case of external galaxies, a linear resolution of order a few hundred parsecs
is typical of synchrotron observations.  
The increase in the uncertainty of the correlation function with $l$ can cause 
serious complications in the analysis of interstellar turbulence using power spectra
of synchrotron fluctuations (the Fourier transforms of the correlation function) 
as suggested by \citet{Lazarian:2012,Lazarian:2016} and \citet{Lee:2016}. 
This problem may not be evident when power spectra are considered because it is
difficult to estimate their statistical accuracy. However, correlation analysis, 
with due attention to the errors, makes the problem evident. 

\section{Discussion}
\label{sect:summary}
We have performed detailed correlation analysis of the random physical fields in
extensive ISM simulations, focusing mainly on the warm gas since it occupies a 
larger part of the volume. Statistical properties of the fluctuations in the gas
properties are strongly non-Gaussian because of widespread filamentary and planar,
small-scale structures. Such features cannot be captured by second-order 
correlation functions (or their equivalent, power spectra) and require other tools
sensitive to all statistical moments of the random field, such as Minkowski
functionals \citep[e.g.,][and references therein]{Wilkin:2007,Makarenko:2015} 
and topological data analysis \citep{Adler:2010,Edelsbrunner:2014}. 
However, careful correlation analysis remains a necessary first step in the
exploration of statistical properties of random fields.

There are two difficulties in correlation analysis (and its equivalent, power 
spectrum analysis) that deserve special attention as they also occur in any 
exploration of either simulated or observational data. Correlation analysis is 
only meaningful when applied to a random distribution. Therefore, random 
fluctuations in physical parameters need to be isolated first by subtracting their
averaged distributions. Averaging is straightforward in infinite domains with
statistically homogeneous fluctuations. However, in reality the domain can contain
only a modest number of correlation volumes, and the mean distributions of
physical variables are not necessarily uniform or describable via a simple trend. 
We obtain the averaged distributions using Gaussian smoothing at a scale 
(half-width of the Gaussian window) of $50\p$ chosen carefully as in
\citet{Gent:2013b} (see Section~\ref{sect:averaging}). 
Simpler procedures, for example using a uniform mean value at a given $z$, distort 
the results because of the contamination of the structure and correlation functions 
by systematic and complicated non-random trends. 
In particular, the values of correlations lengths obtained under the assumption of
horizontally uniform mean values are unphysically large, exceeding $200\p$.

Even with a correlation lengths $l_0$ of less than $100\p$, the finite size of the
domain (of order $1\kpc^3$ in our case) can significantly affect the estimated 
values of $\lo$, as the integration in Equation~\eqref{eq:length} extends to 
infinity. We resolve this problem by fitting the measured correlation functions with
physically motivated forms, which can then be integrated over an infinite range. The
difference between the correlations lengths obtained with and without this fitting 
can be as large as a factor of two.

Given the complex structure of the simulated ISM, it is not surprising that 
different physical variables have different correlation functions and different
correlation lengths $l_0$, as shown in Table~\ref{tab:Table1}. 
The observational estimates available for the correlations lengths in the ISM 
provide a wide range of values depending on the quantity observed. 
Conclusive comparison with observations requires detailed knowledge of the
statistical properties of the random fields involved and their cross-correlations
\citep{Stepanov:2014}. Interstellar turbulence cannot be characterized by a single
correlation length.

We have estimated the correlation time of the velocity fluctuations $\tau_0$. 
In the simulations used here,$\tau_0 \simeq 5 \Myr$ is close to both the eddy 
turnover time, $\tau_\mathrm{eddy} \simeq 8 \Myr$ and the estimated time
interval between the passage of shock fronts through a given position,
$\tau_\mathrm{shock} \simeq 1 \Myr$. The correlation time is likely to be sensitive 
to the supernova rate (and then, star formation rate) and may be closer to
$\tau_\mathrm{shock}$ when the supernova rate is higher.
Further calculations with varying supernova rates are needed to explore under what
conditions either physical process dominates the correlation time.

The random magnetic field is noticeably anisotropic, with larger rms values for
azimuthal ($y$) and vertical ($z$) components in comparison to the radial ($x$)
component, with $b_{z}$ the strongest component.
The enhanced $y$-component is produced by the action of the large-scale velocity 
shear on the radial turbulent magnetic field $b_{x}$, with the enhanced $z$ 
component produced by stretching of the horizontal magnetic field by the random 
part of the vertical velocity $u_{z}$.  
From the rms values of $b_{x}$ and $b_{y}$, we estimate a degree of polarization of 
$p \approx 0.15$ that may be produced by the magnetic anisotropy.

We also performed correlation analysis of the Faraday depth along the 
vertical direction through the computational domain. Its correlation scale, 
$120 \p$, is significantly larger than the correlation scales of electron 
density ($60 \text{--}90 \p$) and of vertical magnetic field ($60 \p$).
This suggests that there is no simple and universal relationship between 
the correlation scales of electron density, vertical magnetic field and Faraday 
depth. 

Analysis of the total and polarized synchrotron intensities is hampered 
by a rapid increase of the scatter of data points around the average 
contributions to the structure and correlation functions. This difficulty
is evident in the correlation analysis but would not be apparent in the power 
spectra, where statistical errors are difficult to estimate.

\section*{Acknowledgements}
  JFH acknowledges financial support from EPSRC (UK) Grant 1515172.
  Financial support from the Academy of Finland Centre of Excellence ReSoLVE 
  (project number 272157) is acknowledged (FAG); 
  Support: Grand Challenge project SNDYN, CSC-IT Center for Science Ltd. (FAG); 
  AS, AF and GRS were supported by the Leverhulme Trust 
  Grant RPG-2014-427 and STFC Grant ST/N000900/1 (Project 2).
  We are grateful to the referee for careful reading of the 
  manuscript and useful suggestions.
 
\appendix

\begin{table*}
\centering
\caption{\label{tab:Table7}
Root-mean-square values and correlation lengths of the random magnetic and velocity fields 
in the standard and larger domains, {for simulations in the kinematic stage of dynamo action}.}
\begin{tabular}{l l c cc c cc}
\hline\hline
 													&\colhead{} &&\multicolumn{2}{c}{rms fluctuations}					&&\multicolumn{2}{c}{$l_0$ [pc]}\\
  																		\cline{4-5} 										\cline{7-8} 																									 
$|z|$ [pc]									&&&$0$										&$400$									&&$0$ 						&$400$\\
\hline
Random Magnetic field [$\mu$G]			& Standard domain	&&$0.196 \pm 0.001$				&$0.166 \pm 0.001$				&&$64 \pm 2$				&$60 \pm 3$\\ 
 										 				& Larger domain	&&$0.161 \pm 0.001$				&$0.123 \pm 0.001$					&&$64 \pm 3$			&$49 \pm 1$\\[3pt] 
Random speed [km\,s$^{-1}$]	& Standard domain	&&$33.4 \pm 0.2$ 				&$16.2 \pm 0.1$ 					&&$83 \pm 4$ 			&$142 \pm 5$\\ 
 														& Larger domain	&&$11.3 \pm 0.1$ 					&$3.12 \pm 0.04$ 							&&$63 \pm 3$ 			&$79 \pm 3$\\
\hline 
\end{tabular} 														
\end{table*}

\begin{figure*}
\centering
\includegraphics[width=0.47\textwidth]{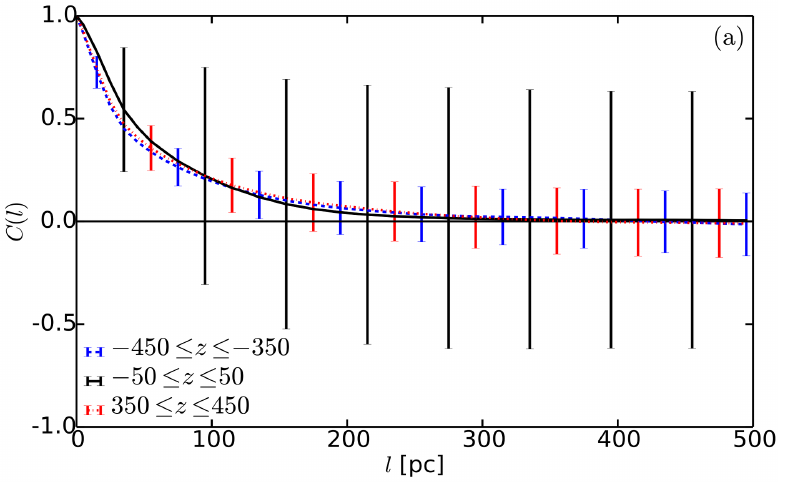}
\hfill
\includegraphics[width=0.47\textwidth]{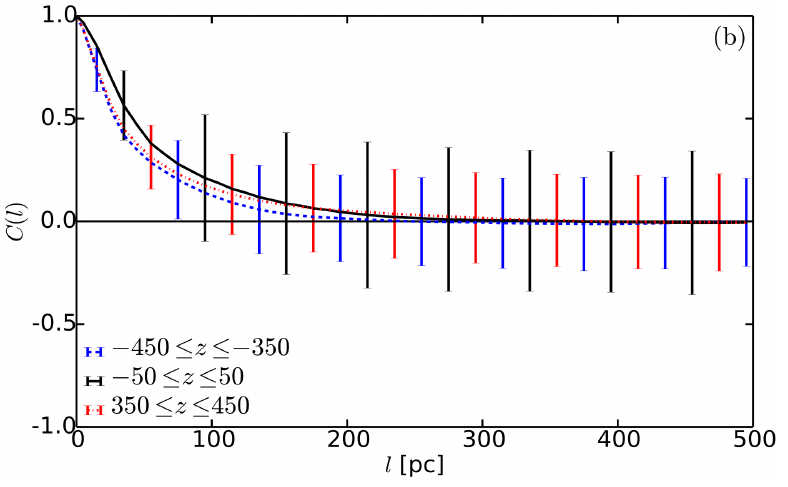}
\caption{Comparison of structure functions for random magnetic field strength $b$, for (a) the standard domain and (b) the larger domain; averaged about $z=-400 \p$ (blue), $0 \, \p$ (black), and $400 \p$ (red). Both plots use data from the kinematic phases of the simulations.
\label{fig:Figure15}
}
\end{figure*}

\begin{figure*}
\centering
\includegraphics[width=0.47\textwidth]{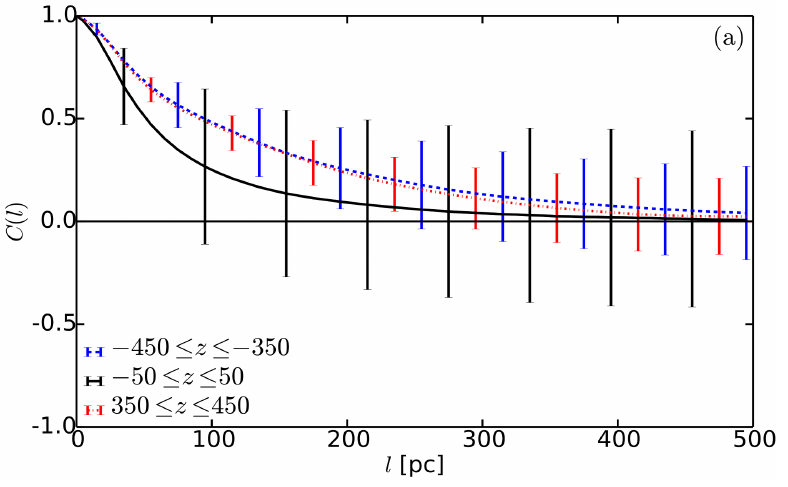}
\hfill
\includegraphics[width=0.47\textwidth]{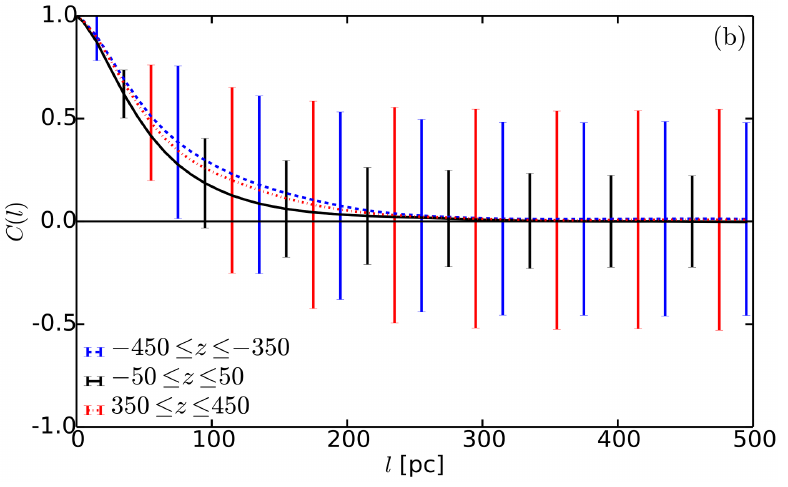}
\caption{Comparison of structure functions for random speed $u'$, comparing domain size as in Figure~\ref{fig:Figure15}.
\label{fig:Figure16}
}
\end{figure*}

\section{Comparison with larger domain}\label{sect:domain_comparisons}

The computational domain used to obtain our results, about $1\times1\times2\kpc^3$,
contains only about $10^3$ correlation cells and, in addition, may be too small to
accommodate the most rapidly growing mode of the large-scale magnetic field. 
The large-scale dynamo remains in its kinematic stage in the larger domain, but
otherwise the simulation has achieved a statistically steady state.
Therefore, we verify the results using similar simulations in a larger domain,
approximately $1.532 \times 1.532 \times 2.556  \, \kpc^{3}$ in size. 
The velocity shear rate is that of the Solar neighborhood, $q=-1$, and
we analyze data from $12$ snapshots in the range $0.336 \leq t \leq 0.6 \Gyr$, 
with a separation of $24 \Myr$.
The extended domain is not designed to capture fountain flows (see 
Section~\ref{sect:parameters}) but is instead employed to to test
how robust our results are to the horizontal area of the simulation.

The results from the larger domain are compared with those obtained from the 
kinematic stage of the large-scale dynamo in the main run discussed in the text. 
We use data from 21 snapshots in the range $0.4\le t \le 0.61 \, \mathrm{Gyr}$, 
with a separation of $10 \, \mathrm{Myr}$.

We find very similar correlations in $b$ between the two runs (see
Figure~\ref{fig:Figure15} and Table~\ref{tab:Table7}), but there are more 
significant differences for $u'$ (see Figure~\ref{fig:Figure16} and
Table~\ref{tab:Table7}; the latter also gives comparable statistics for a similar
kinematic state in the standard domain). 
The correlation lengths of $u'$ are actually smaller for the larger domain, so the
difference does not simply result from velocity structures having been restricted in
size. In light of the differences noted above, further simulations are needed before 
a direct comparison can be made.

\bibliography{references}

\end{document}